\documentclass{aa}
\usepackage{graphicx}
\usepackage[authoryear]{natbib}
\bibpunct[, ]{(}{)}{;}{a}{}{,}
\usepackage{psfig}

\begin{document}

\title{VLT Spectroscopy of Globular Cluster Systems\thanks{Based on
observations collected at the European Southern Observatory, Cerro
Paranal, Chile under programme ID P65.N-0281, P66.B-0068, and
P67.B-0034.}}

\subtitle{II. Spectroscopic Ages, Metallicities, and [$\alpha$/Fe]
Ratios of Globular Clusters in Early-Type Galaxies}

\author{Thomas H. Puzia \inst{1,2,}\thanks{ESA Research Fellow. 
Space Telescope Division of ESA},
 Markus Kissler-Patig \inst{3},
Daniel Thomas\inst{4,5}, Claudia Maraston \inst{4,5}, Roberto P. Saglia
\inst{4}, Ralf Bender \inst{2,4}, Paul Goudfrooij \inst{1}, \& Maren
Hempel \inst{3}}

   \offprints{Thomas H. Puzia, \email{tpuzia@stsci.edu}}

   \institute{
   	Space Telescope Science Institute, 3700 San Martin Drive,
        Baltimore, MD 21218, USA, \\\email{tpuzia, goudfroo@stsci.edu}
      \and
   	Sternwarte der Ludwig-Maximilians-Universit\"at,
        Scheinerstr. 1, 81679 M\"unchen, Germany,
      \and
      European Southern Observatory, 85749 Garching bei M\"unchen,
        Germany, \\\email{mkissler, hempel@eso.org}
      \and
      Max-Planck-Institut f\"ur extraterrestrische Physik,
        Giessenbachstrasse, 85748 Garching bei M\"unchen, Germany,
        \\\email{bender, saglia@mpe.mpg.de}      
      \and
      University of Oxford, Astrophysics, Keble Road, Oxford, OX1 3RH, United Kingdom,
      	\\\email{dthomas, maraston@astro.ox.ac.uk}
        }

   \authorrunning{Puzia et al.}  
   \titlerunning{VLT Spectroscopy of Globular Cluster Systems} 

   \date{Received ... ; accepted ... }

   \abstract{An analysis of ages, metallicities, and [$\alpha$/Fe] ratios of globular
cluster systems in early-type galaxies is presented, based on Lick index
measurements summarized in Puzia et al.~(2004, Paper I of this series). In
the light of calibration and measurement uncertainties, age-metallicity
degeneracy, and the relative dynamic range of Lick indices, as well as
systematics introduced by abundance ratio variations (in particular
variations of [$\alpha$/Fe] ratios), we find that the most reliable age
indicator for our dataset is a combination of the Lick Balmer-line indices
H$\gamma_{\rm A}$, H$\beta$, and H$\delta_{\rm A}$. [MgFe]\arcmin\ is used
as a spectroscopic metallicity indicator which is least affected by
[$\alpha$/Fe] variations. We introduce an interpolation routine to
simultaneously derive ages, metallicities, and [$\alpha$/Fe] ratios from
diagnostic grids constructed from Lick indices. From a comparison of
high-quality data with SSP model predictions, we find that $\sim\!2/3$ of
the globular clusters in early-type galaxies are older than 10 Gyr, up to
$1/3$ have ages in the range $\sim5\!-\!10$ Gyr, and only a few cluster are
younger than $\sim\!5$ Gyr. Our sample of globular clusters covers
metallicities from [Z/H]~$\approx-1.3$ up to $\sim0.5$ dex. We find that
metal-rich globular clusters show on average a smaller mean age and a
larger age scatter than their metal-poor counterparts. [$\alpha$/Fe]
diagnostic plots show that globular cluster systems in early-type galaxies
have super-solar $\alpha$/Fe abundance ratios with a mean
[$\alpha$/Fe]~$=0.47\pm0.06$ dex and a dispersion of $\sim0.3$ dex. We
find evidence for a correlation between [$\alpha$/Fe] and metallicity, in
the sense that more metal-rich clusters exhibit lower $\alpha$-element
enhancements. A discussion of systematics related to the Lick index system
shows that the method suffers to some extent from uncertainties due to
unknown horizontal branch morphologies at high metallicities. However,
these systematics still allow us to make good qualitative statements. A
detailed investigation of indices as a function of data quality reveals
that the scatter in Balmer index values decreases for higher-quality data.
In particular, extremely low Balmer index values that are lower than any
SSP model prediction tend to disappear. Furthermore, we find that observed
photometric colors are in good agreement with computed SSP colors using
ages and metallicities as derived from the spectroscopic line indices.
   }

   \maketitle


\section{Introduction}
\label{ln:intro}
Two prominent models of early-type galaxy formation are lively discussed
in the literature. In the {\it monolithic-collapse} scenario
\citep[][~etc.]{tinsley72, larson75, silk77, arimoto87} the majority of
stars in early-type galaxies forms early, at redshifts $z\ga 2$. Empirical
scaling laws, such as the fundamental plane, the Mg--$\sigma$ relation,
and the [$\alpha$/Fe]-$\sigma$ relation support such an early formation
epoch \citep{bower92, bender96, bender96a, treu99, kuntschner00,
trager00a, trager00b, thomas02, peebles03}. On the other hand, the {\it
hierarchical merging} picture \citep{white78, kauffmann93,
baugh98, cole00, somerville01} sees early-type galaxies as the result of
multiple merging and accretion events of smaller units over an extended
period of time until the very recent past. In this way a significant
fraction of stars is formed below a redshift of unity. Observed ongoing
mergers and accretion events \citep[e.g.][]{whitmore95, ibata95, koo96}
and the existence of kinematically decoupled cores \citep{franx88,
bender88, jed88, surma95, davies01} are clearly enforcing arguments for
this scenario.

These models describe two antipodal paradigms of galaxy formation. The
quantification of their importance as a function of redshift, environment,
and galaxy morphology is necessary to provide a detailed insight in galaxy
formation and assembly. A major difference between the two pictures are
the different star-formation histories of early-type galaxies. While in
the hierarchical picture most massive galaxies are thought to experience
long assembly time scales, the monolithic collapse scenario predicts very
early and short bursts of star formation. Clearly, the predicted
star-formation histories stand in marked contrast and are an important
piece of evidence to differentiate between these two models.

However, to recover star-formation histories, one ideally has to resolve
the underlying stellar populations which build up a galaxy. Only a few
early-type systems are close enough so that their halos can be resolved
into single stars (e.g. in M32: \citealt{grillmair96}, NGC~3115:
\citealt{elson97}, NGC~5128: \citealt{harris02}, NGC~3379:
\citealt{gregg04}). Most photometric and spectroscopic studies of the
unresolved diffuse light are hampered by the mix of ages, metallicities,
and abundance ratios in the stellar populations
\citep[e.g.][]{maraston00}. In combination with the well-known
age-metallicity degeneracy \citep{faber72, oconnell76, worthey94m}, it is
extremely difficult to disentangle even the different major stellar
populations from studies of the diffuse light only, let alone to
reconstruct a detailed star formation history.

\subsection{Globular Cluster Systems as Diagnostics of Star-Formation
Histories} Globular cluster systems are very useful tools to study star
formation histories of galaxies. Several arguments support the hypothesis
that globular clusters trace major star-formation events in galaxies. (1)
The formation of massive star clusters, which are likely to survive a
Hubble time as globular clusters, is observed in interacting/merging and
starburst galaxies \citep[e.g.][]{whitmore95, schweizer97, johnson99,
goudfrooij01, homeier02}. (2) Young massive star clusters are observed in
``simmering'' late-type galaxies. Their number is correlated with the star
formation rate per unit area in these systems \citep{larsen00}. (3)
Normalizing the number of globular clusters to the total baryonic mass of
the host, reveals a surprisingly constant value \citep{mclaughlin99} and
points towards a tight link between star and globular cluster formation.

In other words, globular clusters are fossil records of major
star-formation episodes in galaxies. Their ages, metallicities, and
abundance ratios can provide detailed information on the formation
histories of their host systems. As simple stellar populations,
globular clusters consist of stars characterized by one age and one
metallicity. Hence, the interpretation of their observed colours and
spectra is less ambiguous than for the diffuse light. Their ubiquity in
all galaxy types and their high surface brightness make them easy to
observe out to large distances.

\subsection{Spectroscopy of Globular Cluster Systems}
Spectroscopy opens an independent way to access ages and metallicities of
globular clusters besides photometry. Low resolution spectroscopy
(R~$\la1000$) provides information on the strength of prominent diagnostic
features, such as the Balmer series and some relatively strong iron and
other metal features. Moreover, it can provide clues on the basic
chemistry of globular cluster systems.

The Lick index system standardizes the measurement of spectroscopic line
indices for many strong absorption features \citep{burstein84, worthey94,
worthey97, trager98}. The outstanding role of the Lick system is its
provision of index measurements for many stars with a wide range in $\log
g, T_{\rm eff}$, and [Fe/H]. With this information, so-called fitting
functions\footnote{A fitting function gives the strength of a specific
line index as a function of $\log g, T_{\rm eff}$, and [Fe/H].} are
computed which are an essential ingredient of theore\-tical SSP model
predictions. Using index response functions\footnote{A response function
gives the fractional change of an index as the result of an abundance
change of a given element or the total metallicity.} \citep{tripicco95},
recent SSP models \citep{trager00a, thomas03, thomas04} provide also
information on fundamental abundance ratios, such as [$\alpha$/Fe]. The
combination of age-sensitive (e.g. Balmer-line indices) and
metallicity-sensitive indices ($\langle$Fe$\rangle$, Mgb, [MgFe]\arcmin,
etc.) allows in principle to derive accurate spectroscopic ages and
metallicities. This approach is less affected by the age-metallicity
degeneracy than broadband photometry. In addition, combining indices
sensitive to the abundance of $\alpha$-elements and iron can provide
important clues on star-formation time scales \citep{tinsley79,
matteucci94, greggio97, thomas99}.

The present paper makes use of the spectroscopic data presented in
Puzia et al. (2004, hereafter Paper I) to derive global ages,
metallicities, and [$\alpha$/Fe] ratios for globular clusters in
early-type galaxies. A more detailed study of these global parameters
as a function of internal galaxy properties, such as morphological
type, environment, galaxy mass, etc. will be presented in subsequent
papers of this series.

We describe the selection of high-quality spectroscopic data in
\S\ref{ln:data}. The best combination of Lick indices to achieve most
reliable age and metallicity estimates is discussed in
\S\ref{ln:bestdiagnostic}. In \S\ref{ln:agemet} we derive ages and
metallicities of globular clusters in the studied early-type
galaxies. Global [$\alpha$/Fe] ratios and correlations with age and metallicity
for these globular clusters are presented in \S\ref{ln:alpha}. 
We discuss our results in \S\ref{ln:discussion}.

\section{Selection of Data}
\label{ln:data}
\begin{table*}[ht!]
\centering
\caption[width=\textwidth]{Basic information on host
  galaxies. According to the column the references are: (1)
  \cite{RC3}; (2) \cite{tonry01}; (3) \cite{tully88}; (4) and (5)
  \cite{kissler-patig97}, \cite{ashman98}.}
\label{tab:galdat}
\begin{tabular}{l c c c c c}
\hline\hline
\noalign{\smallskip}
Galaxy   &  type    & $(m-M)_V$    & $M_B$ & $N_{\rm GC}^{\rm a}$ & $S_{\rm N}^{\rm b}$\\
         &   (1)    &  (2)         & (3)   & (4)                  & (5)\\
\noalign{\smallskip}
\hline
\noalign{\smallskip}
NGC~1380 &$-2$/LA   &31.23$\pm0.18$ &$-20.04$ & $560\pm30$   &$1.5\pm0.5$\\
NGC~2434 &$-5$/E0+  &31.67$\pm0.29$ &$-19.48$ & \dots        &\dots      \\
NGC~3115 &$-3$/L$-$ &29.93$\pm0.09$ &$-19.19$ &$520\pm120$   &$1.6\pm0.4$\\
NGC~3379 &$-5$/E1   &30.12$\pm0.11$ &$-19.39$ &$300\pm160$   &$1.2\pm0.6$\\
NGC~3585 &$-5$/E6   &31.51$\pm0.18$ &$-20.93$ & \dots        &\dots      \\
NGC~5846 &$-5$/E0   &31.98$\pm0.20$ &$-21.16$ &$2200\pm1300$ &$3.5\pm2.1$\\
NGC~7192 &$-4.3$/E+ &32.89$\pm0.32$ &$-20.55$ & \dots        &\dots      \\
\noalign{\smallskip}
\hline
\end{tabular}
\begin{list}{}{}
\item[$^{\mathrm{a}}$] Total number of globular clusters
\item[$^{\mathrm{b}}$] Specific frequency, $S_{\rm N}=N_{\rm GC}\cdot
10^{\, 0.4\cdot(M_V + 15)}$ \citep{harris81}
\end{list}
\end{table*}

Our entire sample contains 143 globular clusters in seven early-type
galaxies. About half of our spectra satisfy the ideal S/N standards
($\sim\!30$ per \AA) to derive accurate ages, metallicities, and
[$\alpha$/Fe] ratios. It is possible to achieve an age resolution of
$\sim1$ Gyr (at ages $\sim15$ Gyr) only for the brightest globular
clusters in our sample. The typical separation between the 15 and 14 Gyr
isochrone in current SSP models is of the order
$\Delta$H$\beta\approx0.05$ \AA, and $\sim0.1$ \AA\ for the higher-order
Balmer indices. This separation increases towards younger ages. Hence, the
final sample has to be built from a compromise between age/metallicity
resolution and sample size. We, therefore, set the selection to clusters
with a statistical measurement uncertainty of $\leq0.4$ \AA\ and $\leq0.6$
\AA\ for H$\beta$ and higher-order Balmer line indices, respectively. This
selection corresponds to a minimum age resolution $\Delta t/t\sim0.3$. As
a metallicity indicator, we use the composite [MgFe]\arcmin\
index (see below). An error cut at 0.2 \AA\ for this index guarantees a
metallicity resolution of $\sim0.25$ dex at high and $\sim0.4$ dex at low
metallicities. The above selection criteria leave 71 globular cluster
spectra in our sample which correspond to $\sim$50\% of the initial data.
Henceforth, we consider only these selected globular clusters and combine
the data of all sample galaxies to increase statistical significance in
the discussed relations. Note, that this sample is biased towards the
bright end of the globular cluster luminosity function, i.e.~most massive
clusters. Otherwise, it spans a wide range in globular cluster color and
host galaxy properties (see Tab.~\ref{tab:galdat}).

\begin{figure}[!t]
\centering \includegraphics[width=9cm, bb=10 150 610 700]{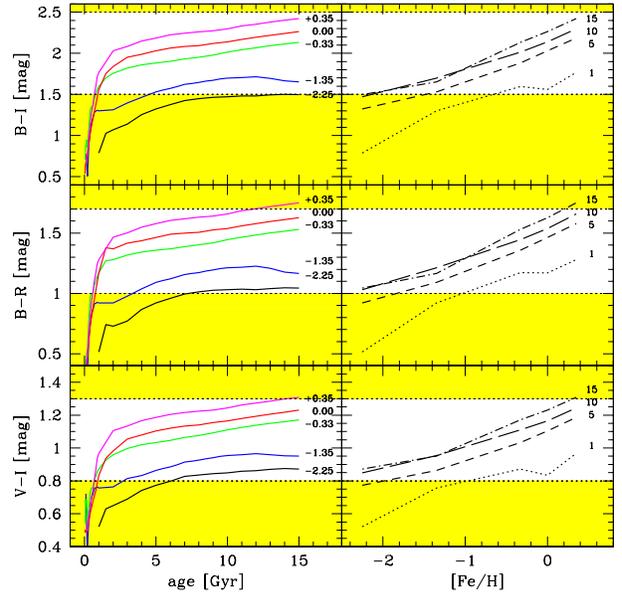}
 \caption{Illustrated are colours which were used to select globular
	cluster candidates as a function of age and metallicity and
	taken from model predictions of \cite{maraston03m}. Upper and
	lower colour cuts are indicated as horizontal lines. Curves in
	left panels are parametrized by metallicity, curves in right
	panels are parametrized by age and indicated accordingly.}
\label{ps:colours}
\end{figure}

Our original colour selection criteria prevented us from observing
globular clusters with a combination of low metallicities [Z/H]~$\la-1.3$
and relatively young ages ($t\la5$ Gyr). However, these objects are
essentially non-existent in our sample galaxies (see Paper I).
Figure~\ref{ps:colours} illustrates the age and metallicity ranges imposed
by our colour selection. All Milky Way globular clusters would be selected
by our colour cuts $0.8\la V-I\la 1.3$, $1.5\la B-I\la 2.5$, and $1.0\la
B-R\la 1.7$ (we refer to Paper I for details). The applied colour cuts
also suggest that metal-rich old globular clusters might be missing in our
final sample. We point out that, within our luminosity cut ($V\la23$ mag),
the fraction of such objects amounts only a few percent. In order to fill
the slit-masks we included cluster candidates redder than our colour cuts.
However, only two genuine globular clusters were found among $\sim\!30$
''mask-filler'' objects. Unlike the expected old ages, we find
intermediate ages (5-10 Gyr) for these two globular clusters. The fraction
of old to intermediate-age, very metal-rich globular clusters is likely to
increase in a sample reaching fainter magnitudes. However, given our
luminosity selection, the number of missed objects is negligible.

Our dataset does not guarantee a strict consistency of the derived mean
ages, metallicities and [$\alpha$/Fe] ratios among the galaxies of our
sample. These parameters are subject to change, since neither the sampled
fraction of each globular cluster system, nor the sampling of colour
distributions and luminosity functions is identical from galaxy to galaxy.
Given the still relatively small numbers of globular clusters per galaxy,
a peculiar age and/or metallicity distribution in one galaxy can influence
the relation between age, metallicity and [$\alpha$/Fe] of the complete
sample, although we verified that no galaxy represents a clear outlier
(see future papers in these series).

\section{Reducing Systematic Uncertainties}
\label{ln:bestdiagnostic}
In the following we determine the best combination of indices as
diagnostics for age and metallicity. Taking into account the uncertainty
of our line index measurements, the mean uncertainties of the Lick system,
and the limits on the predicting power of SSP models, we construct the
relatively best combination of diagnostic diagrams from Lick line indices.
The found combination maximally reduces the internal uncertainties of the
Lick system and the age-metallicity degeneracy of line indices as well as
the influence of abundance ratio variations.

\subsection{Influence of the Blue Horizontal Branch at High Metallicities}
\label{ln:bhb}
As shown by \cite{maraston00} for H$\beta$ and by \cite{maraston03} for
the higher-order Balmer lines, the morphology of the Horizontal Branch
(HB), when extended to warm temperatures ($\sim10000$ K), plays a major
role in increasing the strength of Balmer indices. This effect is due to
metallicity and confuses the use of Balmer lines as pure age indicators
(on this topic see also \citealt{defreitas95} and \citealt{lee00}). Since
the HB morphology cannot be predicted by first principles of stellar
evolution, as it is determined by mass-loss, the line indices need to be
calibrated with globular clusters for which the HB morphology is known
\citep{maraston00, maraston03}. This exercise is clearly impossible for
extragalactic globular clusters and is the reason why in the following we
use both models by \cite{maraston03} which include red and blue HB
morphologies as a function of metallicity. These models encompass the
observed range of Balmer lines in Milky Way globular clusters. As
thoroughly explored by \cite{greggio90}, blue HBs are in principle
possible also in metal-rich ($Z\ga Z_\odot$) stellar populations
\citep[see also][for examples of two metal-rich Milky Way globular
clusters]{rich97}, that suffer from enhanced mass-loss and/or have a high
Helium abundance. Models with blue HBs at high metallicity are published
elsewhere \citep{maraston03m}, and here we report on their differential
effect.

\begin{figure}[!hb]
\centering 
\includegraphics[width=8.5cm, bb=10 150 610 700]{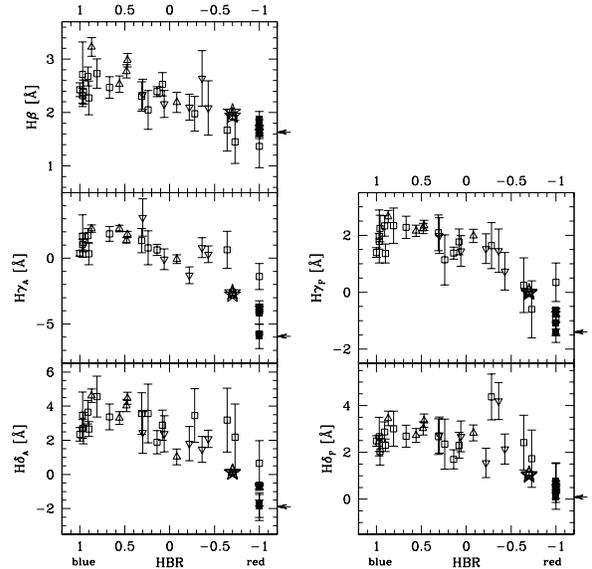}
\caption{The behavior of Lick Balmer indices as a function the horizontal branch
morphology (HBR). This parameter is defined in \cite{lee94} as HBR
(B$-$R)/(B+V+R), where B and R are the number of stars blue-wards and red
wards of the instability strip. V is the number of variable stars inside
the instability strip. HBR~$=1$ indicates an entirely blue and HBR~$=-1$
and entirely red horizontal branch. Here we plot data for globular
clusters in the Milky Way ({\it squares}: data from \citealt{puzia02} and
\citealt{trager98}), M31 ({\it inverted triangles}: \citealt{puzia04},
\citealt{rich03}, and \citealt{trager98}), and the Large Magellanic Cloud
{\it triangles}: \citealt{beasley02}). Filled symbols show globular
clusters with a metallicity [Z/H]~$ -0.6$ (note, all have HBR~=$-1$).
NGC~6388 and NGC~6441 are indicated by stars at HBR~$=-0.7$. A small arrow
at each panel's right ordinate indicates the most extreme Balmer index
value for globular clusters with [Z/H]~$> -0.6$. Several clusters show a
tendency for higher Balmer indices at HBR~$\approx -0.5$. Their deviations
from the overall sequence are $\leq 2\sigma$.}
\label{ps:balmer_hbr}
\end{figure}

To empirically estimate the effect of a varying horizontal branch
morphology on Balmer indices at high metallicities, which might be
responsible for the increased Balmer indices of metal-rich globular
clusters, we parametrize the HB morphology with the HBR parameter
\citep{lee94}. Figure~\ref{ps:balmer_hbr} shows the strength of Balmer
line indices as a function of HB morphology (HBR). In general, blue
horizontal branches produce significantly stronger Balmer indices for
globular clusters in the Milky Way \citep{trager98, puzia02},
M31\footnote{We derive HBR parameters for globular clusters in M31 from
colour-magnitude diagrams using HST data kindly provided by Michael Rich
(GO:6671).} \citep{trager98, puzia04}, and the LMC \citep{beasley02}. Each
panel shows a sequence of metallicity, where the scatter can be
attributed to the ''second parameter''. We use the globular clusters
NGC~6388 and NGC~6441 (which host the bluest horizontal branches among
metal-rich Galactic globular clusters, indicated by stars in
Fig.~\ref{ps:balmer_hbr}) and clusters at [Z/H]~$\ga -0.6$ (filled
symbols) which have entirely red HBs (e.g. NGC~6356 and NGC~6637) to
derive a representative ``second-parameter'' variation of Balmer line
indices at high metallicities. As this approach is fully empirical and
based on the largest HB morphology fluctuation locally observed, we cannot
rule out that even more extreme HB morphologies for globular clusters at a
given metallicity exist outside the Local Group. We find, in the extreme
case, offsets of 0.4 \AA\ in H$\beta$, 3.3 \AA\ in H$\gamma_A$, 1.4 \AA\
in H$\gamma_F$, 2.0 \AA\ in H$\delta_A$, and 1.0 \AA\ in H$\delta_F$
between metal-rich globular cluster with the {\it weakest} Balmer index
(with entirely red HBs) and NGC~6388 and 6441. The HB morphology has
negligible effect on the [MgFe]\arcmin\ index. Consequently, increased
Balmer indices at high metallicities might well be, at least partly, the
result of HB morphology variations.

In the following we use the Balmer indices as age indicators with
confidence at low metallicity ([Z/H]~$\la-0.6$), because HB morphology is
included in our SSP models and under control. At high metallicity
([Z/H]~$\ga-0.6$), we have the warning in mind that ages could be
degenerate with the presence of unresolved blue HBs. For this reason we
will refer to such young ages as ``formal''.

\subsection{The Influence of [$\alpha$/Fe] Variations on Iso-Age Tracks}
\label{ln:balmeralpha}

\begin{figure*}[!ht]
\centering 
\includegraphics[width=12cm, bb=10 150 600 720]{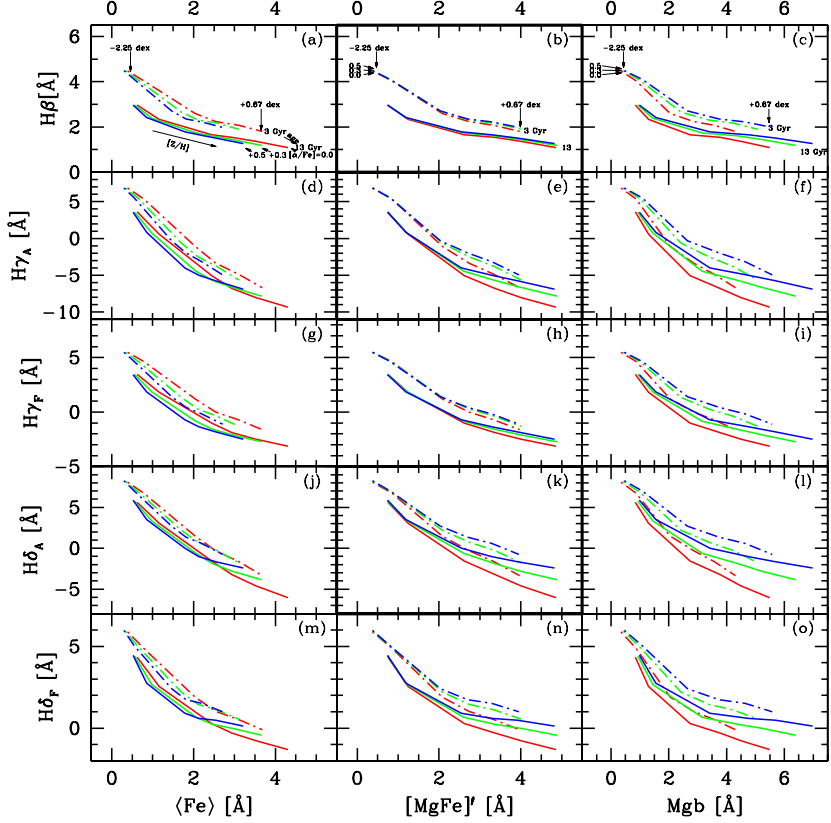}
  \caption{Influence of [$\alpha$/Fe] variations on age-metallicity diagnostic grids.
Panel (a): H$\beta$ vs. $\langle$Fe$\rangle$. Lines show model predictions
\citep{thomas03, thomas04} for stellar populations with metallicities
[Z/H]=$-2.25$ to 0.67 dex and two ages 13 (solid lines) and 3 Gyr
(dot-dashed lines) parametrized for three different [$\alpha$/Fe] ratios
0.0, 0.3, and 0.5 dex. Panel (b) shows the same model predictions for
H$\beta$ vs. [MgFe]\arcmin, while panel (c) illustrates the changes in the
H$\beta$ vs. Mgb grid. The other panels show the predicted variations for
higher-order Balmer line age/metallicity diagnostic grids.}
\label{ps:Balmer_diffmetals}
\end{figure*}

Recently, \cite{thomas03, thomas04} calculated new theoretical Lick index
predictions which are parametrized for {\it well-defined} [$\alpha$/Fe]
ratios for a wide range of ages and metallicities. These models take into
account the effects of changing element abundance ratios on Lick indices,
hence give Lick indices not only as a function of age and metallicity, but
also as a function of the [$\alpha$/Fe] ratio. They are based on the
evolutionary population synthesis code of \cite{maraston98}. The impact
from element-ratio changes is computed with the help of the
\cite{tripicco95} and \cite{korn04} response functions, using an extension
of the method introduced by \cite{trager00a}. Because of the inclusion of
element-ratio effects, the influence of [$\alpha$/Fe] on Balmer indices
can be studied, and is illustrated in Figure~\ref{ps:Balmer_diffmetals}.
Here we show the influence of [$\alpha$/Fe] variations on frequently used
age/metallicity diagnostic diagrams. In general, variations of Balmer-line
indices for isochrones with [$\alpha$/Fe] ratios between solar and $+0.5$
dex are of the order $\sim0.007-0.25$ \AA\ for low, and $\sim0.18-3.63$
\AA\ for high metallicities, and introduce a relative age uncertainty in
the range $\Delta t/t\sim0.1-0.2$. These variations are due to different
contaminations by metal absorption features inside the Balmer index
passband definitions, as illustrated in Figure~\ref{ps:balmerpassbands}.
The Figure shows the higher-order Balmer indices generally include more
metal absorption lines in their feature and pseudo-continuum passbands
than the H$\beta$ index, which is the Balmer index with narrowest passband
definitions. Hence, among all Lick Balmer line indices, the least
[$\alpha$/Fe]-sensitive index is H$\beta$, followed by H$\gamma_{\rm F}$,
H$\gamma_{\rm A}$, H$\delta_{\rm A}$, H$\delta_{\rm F}$ \citep[see
also][]{thomas03, thomas04}. This exercise demonstrates that at high
metallicities the impact of [$\alpha$/Fe] variations on age/metallicity
determinations (in particular, those of early-type galaxies) can
significantly alter the results and/or introduce spurious correlations.

\subsection{The relatively best Age Indicator}
\label{ln:balmer}
\begin{figure*}[!ht]
\centering 
\includegraphics[width=12.0cm, bb=10 150 610 700]{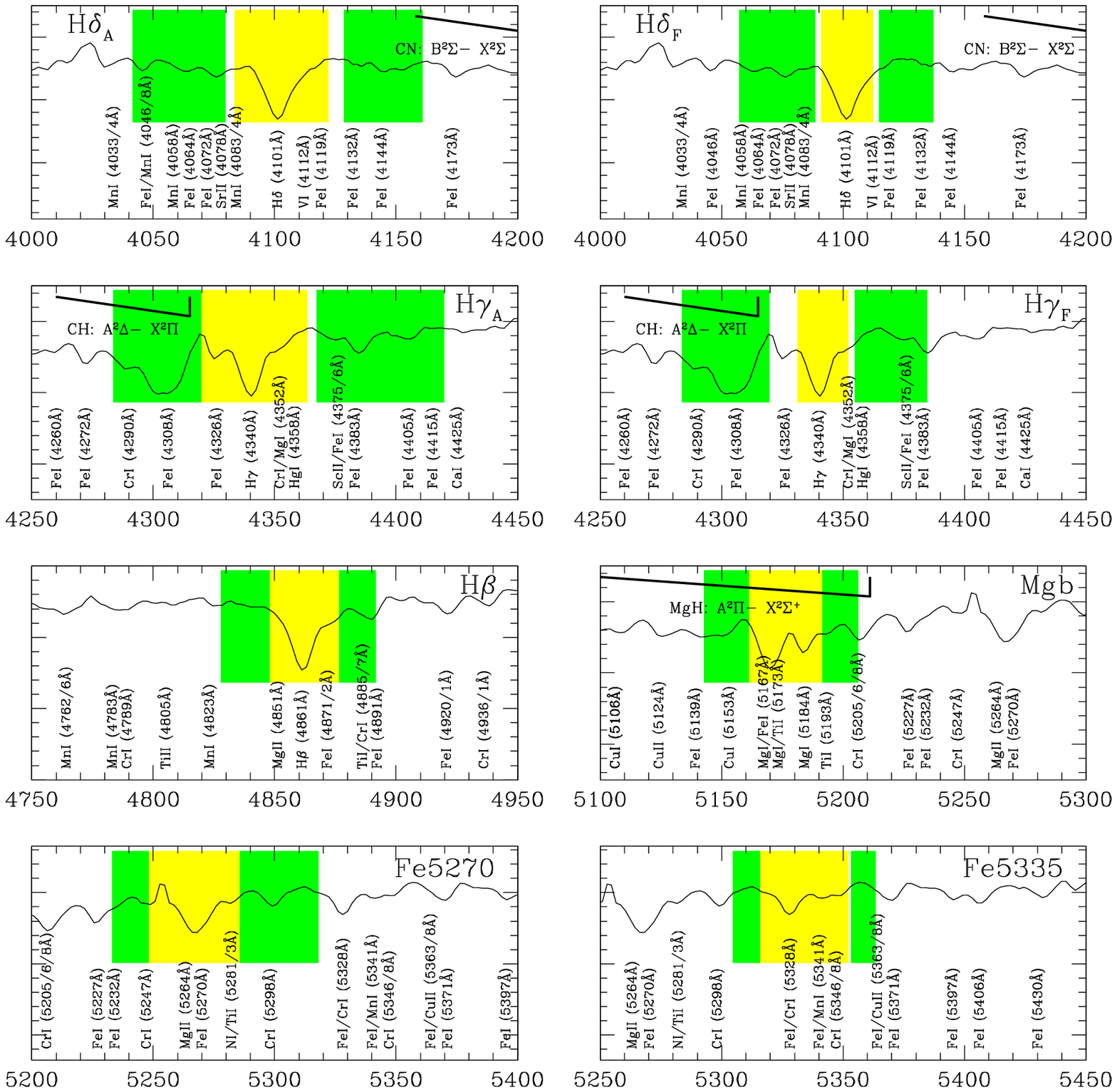}
\includegraphics[width=12.0cm, bb=10 170 610 280]{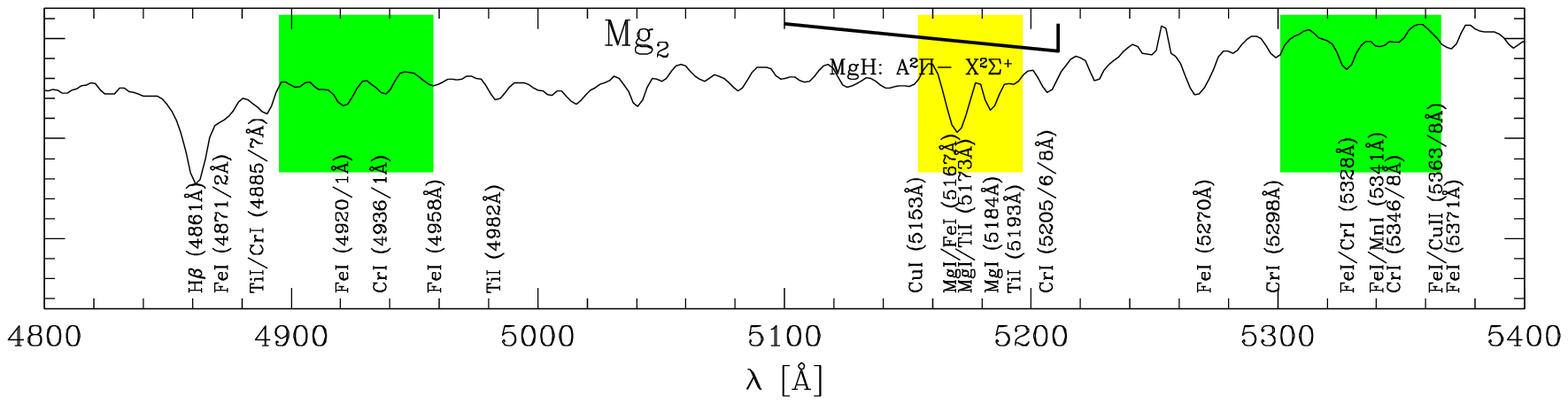}
    \caption{Passband definitions for Balmer-line, Mgb, Fe5270, and Fe5335 Lick indices
with their feature and adjacent continuum passbands. The over-plotted
spectrum is a high-S/N spectrum of the Galactic globular cluster NGC~6284
\citep{puzia02}. The resolution is $\sim7$ \AA\ and was left untouched to
keep satellite lines visible. Line data were taken from \cite{reader81},
data for molecular bands are from \cite{rosen52}. Note the large number of
satellite lines which are included in the passband definitions.}
\label{ps:balmerpassbands}
\end{figure*}

The Balmer line series provides the best spectroscopic age indicator
among the set of Lick line indices. The Lick system defines five
indices (H$\beta$, H$\gamma_{\rm A}$, H$\delta_{\rm A}$, H$\gamma_{\rm
F}$, and H$\delta_{\rm F}$) for three Balmer lines \citep{worthey94,
worthey97}. Figure~\ref{ps:balmerpassbands} shows the passband
definitions for all Balmer indices. In combination with a
metallici\-ty diagnostic, these higher-order Balmer line indices are
widely used to determine (luminosity-weighted) ages and metallicities
of galaxies \citep[e.g.][]{trager98, trager00a, trager00b,
kuntschner00, poggianti01, kuntschner02a, thomas03}.

However, different types of diagnostic plots based on different
Balmer-line indices are used throughout the literature. Although the age
predicting power of an arbitrarily chosen diagnostic plot (most common
versions include the H$\beta$ and Mg$_2$ or $\langle$Fe$\rangle$ indices
might yield accurate-enough results for a specific scientific goal (e.g.
the mean age difference between two different galaxy samples), the choice
of a specific diagnostic plot is still subject to observational
constraints and individual assessment, and makes comparisons between
studies difficult. As a consequence, most authors use several diagnostic
plots with different Balmer line indices and assign equal importance to
the results derived from each of those.

In the following we provide a recipe to define a quantity from which the
relatively best Balmer-line age indicator can be determined. This quantity
takes into account the quality of a given dataset and the diagnostic power
of theoretical predictions from which one intends to derive the age and
metallicity. It does not take into account {\it systematic} uncertainties
in the fitting functions for a particular index.

In particular, the age sensitivity of an index is a function of the
following parameters:
\begin{list}{$\bullet$}{}
\item $\eta$: mean error of the data in \AA
\item $\zeta$: transformation accuracy to the Lick system in \AA
\item $\gamma$: mean error of the original Lick spectra in \AA
\item $\delta$: accuracy of the Lick fitting functions in \AA
\citep{worthey94, worthey97}
\item ${\cal D}_{\rm Z}$: index range in \AA\ covering all ages at a given
metallicity, hereafter termed the dynamic range 
\item ${\cal S}_{\rm \alpha,Z, t}$: degeneracy parameter, which quantifies
the sensitivity of an index $I$ to age, metallicity, and $\alpha$/Fe ratio at a
given [$\alpha$/Fe], metallicity, and age (i.e. the impact of the
age-metallicity and age-$\alpha$/Fe degeneracy), given in dex/Gyr
\end{list}
The numerical values for each parameter are given in
Table~\ref{tab:abcdBalmer} for each Balmer index. It is worth noting that
some of these values are only valid for our data quality in combination
with the SSP models of \cite{thomas03, thomas04}. For different data and
SSP models, $\eta$, ${\cal D}_{\rm Z}$, and ${\cal S}_{\rm \alpha,Z,t}$
are subject to change. To quantify the most age-sensitive and least
metallicity-sensitive Balmer index, we define the quantity
\begin{equation}
\label{eq:R}
{\cal R} = \frac{{\cal D}_{\rm Z}\cdot {\cal S}_{\rm
\alpha,Z,t}}{\sqrt{\eta^2 + \zeta^2 + \gamma^2 + \delta^2}}
\end{equation}
where the degeneracy parameter, ${\cal S}_{\rm \alpha,Z,t}$, is defined as
\begin{equation}
\label{eq:S}
{\cal S}_{\rm \alpha,Z,t}(I) = \left| \frac{\partial I}{\partial t}
\right|_{\rm Z,\alpha} \cdot 
\left(
\left| \frac{\partial I}{\partial{\rm [Z/H]}}\right|^2_{\rm \alpha,t} + 
\left|\frac{\partial I}{\partial{\rm [\alpha/Fe]}}\right|^2_{\rm Z,t}\right)^{-\frac{1}{2}}
\end{equation}

$\cal R$, which is given in dex$\cdot$Gyr$^{-1}$, is essentially the
dynamic range of an index $I$ at a given age and metallicity expressed in
units of the total uncertainty. We use the mean dynamic age range ${\cal
D}_{\rm Z}$ at two different metallicities [Z/H]$=-1.35$ and 0.0 between
the 1 and 15 Gyr isochrone. Each SSP model provides a well-defined
relative age scale which we use here to parametrize ${\cal S}_{\rm
\alpha,Z,t}$ for two different metallicities [Z/H]$=-1.35$ and 0.0, two
different ages 3 and 13 Gyr, and [$\alpha$/Fe]~$=0.3$. ${\cal S}_{\rm
\alpha,Z,t}$ is the ratio of age, metallicity, and [$\alpha$/Fe] partial
derivatives (see Eqn.~\ref{eq:S}). In other words, ${\cal S}_{\rm
\alpha,Z,t}$ is a measure of the age-metallicity and age-$\alpha$/Fe
degeneracy, and is maximal for indices which are most sensitive to age and
least sensitive to metallicity and [$\alpha$/Fe] variations, at the same
time.

The highest ${\cal R}$ indicates the best age indicator with least
age-metallicity and age-$\alpha$/Fe degeneracy. In
Table~\ref{tab:abcdBalmer} we provide values for ${\cal D}_{\rm Z}$ at two
different metallicities and for ${\cal S}_{\rm \alpha,Z,t}$ at
[$\alpha$/Fe]~$=0.3$ and four age-metallicity combinations for each Balmer
line index. Since SSP models do not provide continuous but discrete
predictions the partial derivatives are substituted by difference ratios,
e.g. $\partial I/\partial t \rightarrow\ \Delta I/\Delta t$. The quotients
are determined by linear interpolation of SSP models.

We determine the relatively best age indicator from the set of five Lick
Balmer indices by combining the mean dynamic range $\langle{\cal D}_{\rm
Z}\rangle$, the mean age-metallicity and age-$\alpha$/Fe degeneracy
parameter $\langle{\cal S}_{\rm \alpha,Z,t}\rangle$, and the total index
uncertainty which is the denominator in equation \ref{eq:R}. The final
mean ${\cal R}$ is documented in the last column of
Table~\ref{tab:abcdBalmer}. We find that the relatively best age
diagnostic for our data is the H$\gamma_{\rm A}$ index followed by the
indices H$\beta$ and H$\delta_A$. H$\gamma_F$ and H$\delta_F$ have the
smallest ${\cal R}$ values and are not considered to be reliable age
indicators.

It is instructive to see that despite the relatively large age-metallicity
degeneracy of the H$\gamma_{\rm A}$ index, it still provides the most
accurate age predictions. This fact is primarily due to the large dynamic
range of H$\gamma_{\rm A}$ compared to its mean measurement uncertainty.
H$\beta$, on the other hand, has a relatively large total uncertainty and
the measurements will therefore be more scattered over the diagnostic
plot's parameter range. However, H$\beta$ shows by far the highest
$\langle{\cal S}_{\rm \alpha,Z,t}\rangle$ value (column 13 in
Tab.~\ref{tab:abcdBalmer}; see also Fig.~\ref{ps:Balmer_diffmetals}) and
is therefore least sensitive to the cumulative effects of age-metallicity
and age-$\alpha$/Fe degeneracy. In general, the higher-order Balmer lines
require less S/N to guarantee a similar total index accuracy as H$\beta$,
which is primarily due to the narrow passband definition of H$\beta$ (see
Fig.~\ref{ps:balmerpassbands}). If our dataset would be infinitely
accurate (i.e. $\eta=0$ in Eqn.~\ref{eq:R}), the order of ${\cal R}$ from
the best to worst Balmer index would remain unchanged. The value of ${\cal
R}$ is predominantly governed by uncertainties in the fitting functions of
the respective index. To vary this order, the mean measurement
uncertainties have to be very discrepant and the SSP-model predictions
have to deviate significantly from the model used here. It is expected
that the relative accuracy of Balmer index measurements is comparable
between different datasets as they are usually derived from one optical
spectrum. The {\it relative} age scale of SSP models appears to be quite
stable against the choice of different stellar evolutionary tracks for
ages $>1$ Gyr \citep{charlot96, trager00a, maraston03proc}. This scale is
used in our above prescription. It can therefore be expected that no large
fluctuation in ${\cal R}$ will arise from the use of different SSP model
predictions\footnote{We note that systematic uncertainties in the fitting
functions and effects of emission-line filling are not considered by this
exercise.}.

\begin{table*}[!ht]
\begin{center}
 \caption{Summary of the coefficients relevant to equation
	\ref{eq:R}. The coefficients in columns $2-8$ are given in units
	of \AA . Columns $9-14$ are given in dex/Gyr. Note that 
	for the higher-order Balmer indices column 5 gives the arithmetic
	mean of the uncertainties for cool and warm stars 
	\citep[see Table 3 and 4 in][]{worthey97}.}
\label{tab:abcdBalmer}
{\scriptsize
\begin{tabular}[angle=0,width=\textwidth]{lrrrcccccccccc}
\hline\hline
\noalign{\smallskip}
 index & $\eta$ & $\zeta$ & $\gamma$ & $\delta$ & ${\cal D}_{-1.35}$ &
${\cal D}_{0.0}$ & $\langle{\cal D}_{\rm Z}\rangle$ & ${\cal S}_{0.3,-1.35,3}$ &
${\cal S}_{0.3,-1.35,13}$ & ${\cal S}_{0.3,0.0,3}$ & ${\cal S}_{0.3,0.0,13}$ &
$\langle{\cal S_{\rm \alpha,Z,t}}\rangle$ & $\cal R$ \\  
\noalign{\smallskip}
1 & 2 & 3 & 4 & 5 & 6 & 7 & 8 & 9 & 10 & 11 & 12 & 13 & 14\\
\noalign{\smallskip}
\hline
\noalign{\smallskip}
H$\beta$        &  0.20 & 0.232 & 0.22 & 1.30 & 2.55 & 2.68 & 2.62 & 0.382 & 0.137 & 0.408 & 0.102 & 0.257 & 0.497 \\
H$\gamma_A$&  0.28 & 0.722 & 0.48 & 1.78 & 7.75 & 11.01& 9.38 & 0.283 & 0.068 & 0.227 & 0.036 & 0.154 & 0.722 \\
H$\gamma_F$&  0.28 & 0.448 & 0.33 & 1.34 & 4.55 & 5.83 & 5.19 & 0.291 & 0.072 & 0.123 & 0.050 & 0.134 & 0.471 \\
H$\delta_A$   &  0.27 & 1.043 & 0.64 & 1.27 & 5.73 & 9.11 & 7.42 & 0.259 & 0.045 & 0.180 & 0.032 & 0.129 & 0.537 \\
H$\delta_F$   &  0.28 & 0.790 & 0.40 & 1.18 & 3.83 & 4.47 & 4.15 & 0.281 & 0.052 & 0.114 & 0.036 & 0.121 & 0.334 \\
\noalign{\smallskip}
\hline
\end{tabular}
}
\end{center}
\end{table*}

\subsection{The relatively best Metallicity Indicator}
\label{ln:metals}
The index with the highest metallicity sensitivity and minimal
age-sensitivity could in principle be found in a comparable way as it
was done for the relatively best age diagnostic. The major impact of
typical metallicity tracers, such as $\langle{\rm Fe}\rangle$, Mg$_2$,
and Mg$b$, on the absolute metallicity scale is expected to arise from
changing abundance ratios. To reduce the influence of [$\alpha$/Fe]
variations on age and metallicity determinations, \cite{thomas03}
modify the [MgFe] index\footnote{[MgFe]~$=\sqrt{{\rm
Mg}b\cdot\langle{\rm Fe}\rangle}$, see \cite{gonzalez93}.} to obtain
an [$\alpha$/Fe]-{\it insensitive} metallicity index,
\[ {\rm [MgFe]\arcmin } = \sqrt{{\rm Mg}b\cdot (0.72\times{\rm
Fe5270} + 0.28\times{\rm Fe5335})}.\] 

Figure~\ref{ps:Balmer_diffmetals} shows the behavior of isochrones in
three different frequently used diagnostic plots. The middle column
impressively illustrates that [MgFe]\arcmin, indeed, is essentially
independent of [$\alpha$/Fe]. Henceforth, we adopt the [MgFe]\arcmin\
index as the best metallicity indicator and use it in combination
with selected age indicators (see Sect.~\ref{ln:method}) to derive ages, 
metallicities, and [$\alpha$/Fe] ratios for our sample globular clusters.

\subsection{Excluding contamination by Ionized Gas}
\label{ln:emission}
It is known that $\sim40-60$\% of early-type galaxies show indications of
emission in their absorption spectra \citep{caldwell84, phillips86,
gonzalez93, goudfrooij94}. In a narrow-band imaging survey,
\cite{macchetto96} find ionized gas in $\sim80$\% of early-type galaxies
including flocculent H$\alpha+$[N{\sc ii}] emission in NGC~3379, NGC~5846,
and NGC~7192, well within $\sim1\, R_{\rm eff}$, but no significant
emission in NGC~3115. This gas is located in the central parts and
distributed in a rather regular way, suggestive of a disk. If dominant,
all Balmer indices would be affected, along with potential contamination
of Fe5015 by [O{\sc iii}] ($\lambda$ 5007 \AA) and of Mg$b$ by [N{\sc i}
($\lambda$ 5199 \AA) \citep{goudfrooij96}. In that case, our measurements
would indicate {\it too old ages}. This effect rapidly decreases from
H$\beta$ towards H$\gamma$ and H$\delta$ \citep{osterbrook89},
i.e.~higher-order Balmer indices are less affected and should be
preferentially used for age determinations in the presence of ionized gas.

In order to exclude a major effect of ionized gas on line-strength
measurements in our globular cluster data, we performed several
tests/estimates.

Since line emission is concentrated in the central parts of galaxies
\citep{macchetto96}, we expect a correlation of Balmer indices with
galacto-centric radius if line-emission contamination is significant. We
find no evidence that Balmer indices are correlated with galacto-centric
distance. A more detailed analysis of background spectra shows that most
clusters are located within $\sim2-3\, R_{\rm eff}$ and that the flux
level of the diffuse galaxy light is well below the object flux. In
particular, we find no correlation inside one effective radius, where line
emission is expected to be strongest. Furthermore, we find no correlations
of Balmer indices for globular clusters with Balmer indices measured on
corresponding background spectra.

Visual re-inspection of the background subtraction process for some
low-H$\beta$ outliers (see Fig.~\ref{ps:Balmer_Metal}) underline the good
quality of background modeling and subtraction. However, problems with
accurate background subtraction might occur in very few cases when line
emission has a very filamentary structure. For instance, the worst case
scenario would be when a globular cluster overlaps with a filament of
ionized gas while the slit is aligned perpendicular to such a filament.
However, a filamentary emission pattern is not found in the
\cite{macchetto96} study for our host galaxies. We conclude that line
emission has no measurable effect on the scatter in the age/metallicity
and [$\alpha$/Fe] diagnostic plots.

\section{Ages and Metallicities}
\label{ln:agemet}
\subsection{Iterative Method}
\label{ln:method}
Having determined the relatively best metallicity diagnostic, we determine
the best combination of Balmer-line indices as our prime age indicator. In
deciding whether a Balmer index will be chosen as part of this most
reliable age proxy, we carefully analyze the following points for each
Balmer index and assign priorities in descending order:
\begin{description}
\item[a)] ranking of the reliability parameter ${\cal R}$, as described in Section~\ref{ln:balmer}
\item[b)] including the maximum number of the Balmer lines without duplicate measurements 
by two or more line indices
\end{description}
We choose the combination of H$\gamma_{\rm A}$, H$\beta$, and
H$\delta_{\rm A}$ as our most reliable age indicator. Consequently, ages
and metallicities for individual globular clusters are computed as the
weighted mean of the parameters derived from diagnostic plots constructed
from these three Balmer indices vs. [MgFe]\arcmin. As weights we choose
the reliability parameters ${\cal R}$ from Table~\ref{tab:abcdBalmer}.

Since the SSP model predictions allow full control over [$\alpha$/Fe]
variations within the diagnostic grids, we use an iterative approach in
combination with a second diagnostic grid from which we derive the
[$\alpha$/Fe] ratios (Mg$_{2}$ vs. $\langle$Fe$\rangle$, see
Sect.~\ref{ln:alpha}). As a first step the $\alpha$-enhancement for each
individual globular cluster is derived. This value is used to interpolate
the age/metallicity diagnostic grid for the correct $\alpha$-element
enhancement. From so adjusted grids ages and metallicities are computed
using linear\footnote{Note that the SSP model grids are interpolated
in linearly.} interpolation of the model predictions
employing a least-square technique. These ages and metallicities are then
used to adjust the [$\alpha$/Fe] diagnostic grid, which itself is slightly
dependent on age. This procedure is iterated until the age, metallicity,
and [$\alpha$/Fe] values converge. Extensive tests show that this goal is
reached within a few iteration steps. Along with these values we also
determine the statistical errors due to index measurement uncertainties.
Typically the errors are asymmetric which is a result of the skewness of
diagnostic grids (see Fig.~\ref{ps:Balmer_Metal}).

For objects which fall outside the diagnostic grid, we assign the most
likely extreme grid value and do {\it not} extrapolate the model
predictions, i.e. we project the data onto the closest grid point along
its error vector\footnote{To illustrate the effect of this procedure we
mark objects, which fall off the grid in all diagnostic plots, by
double-hatched histogram in the distributions plots in
Figure~\ref{ps:amah}.}. Note that due to the increasing influence of hot
blue horizontal-branch stars, the iso-age tracks for old ($t>8$ Gyr)
stellar populations with metallicities below $-1.0$ dex tend to overlap.
This introduces an ambiguity in assigning ages and metallicities to
individual globular clusters and artificially broadens the age
distributions at old ages. However, only six objects are affected by this
ambiguity which has no effect on the following results. For these six
objects the routine randomly assigns either young or old ages and the
corresponding metallicities.

\subsection{Consistency Checks}
\label{ln:conschecks}
\begin{figure*}[!th]
\centering 
\includegraphics[width=5.5cm,]{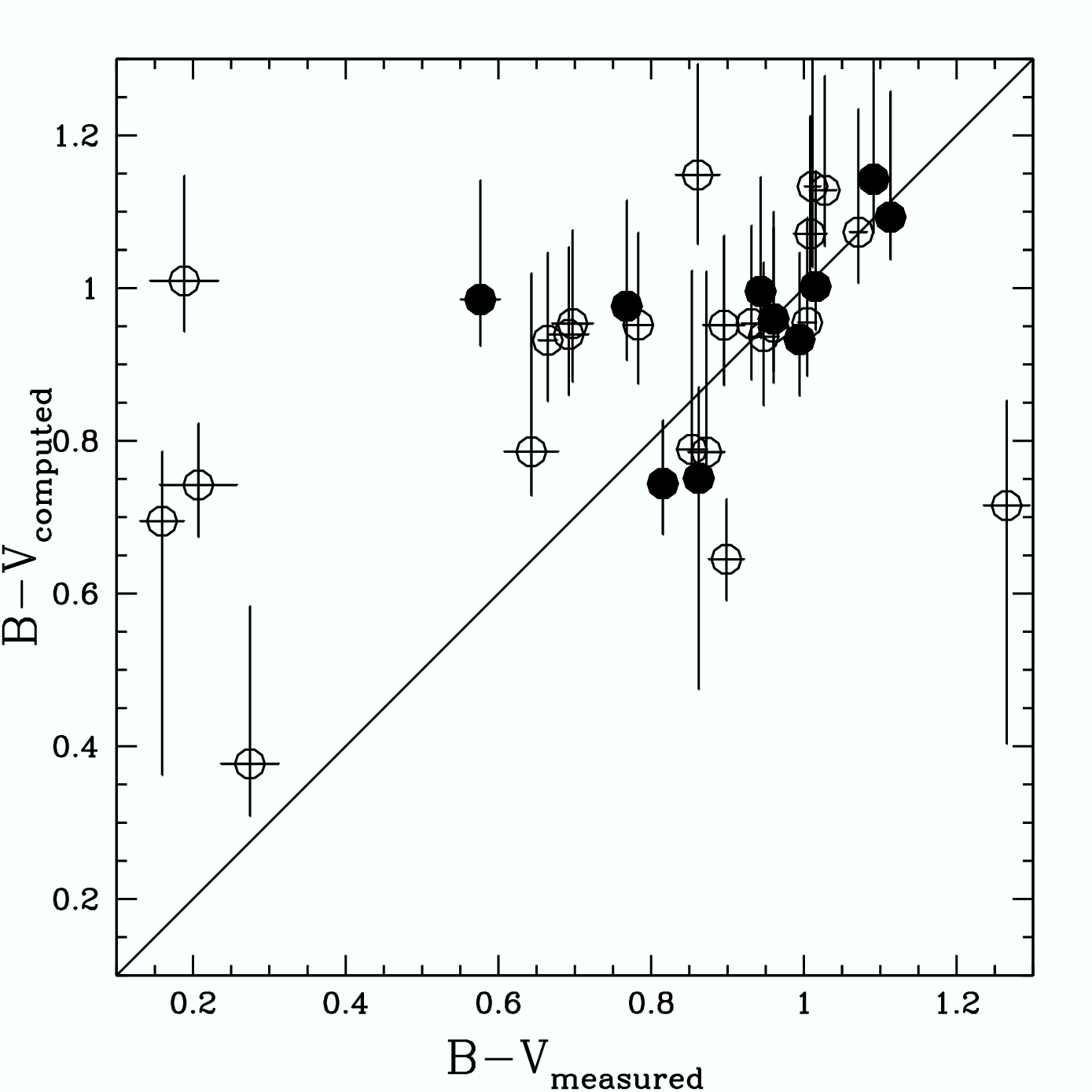}
\includegraphics[width=5.5cm,]{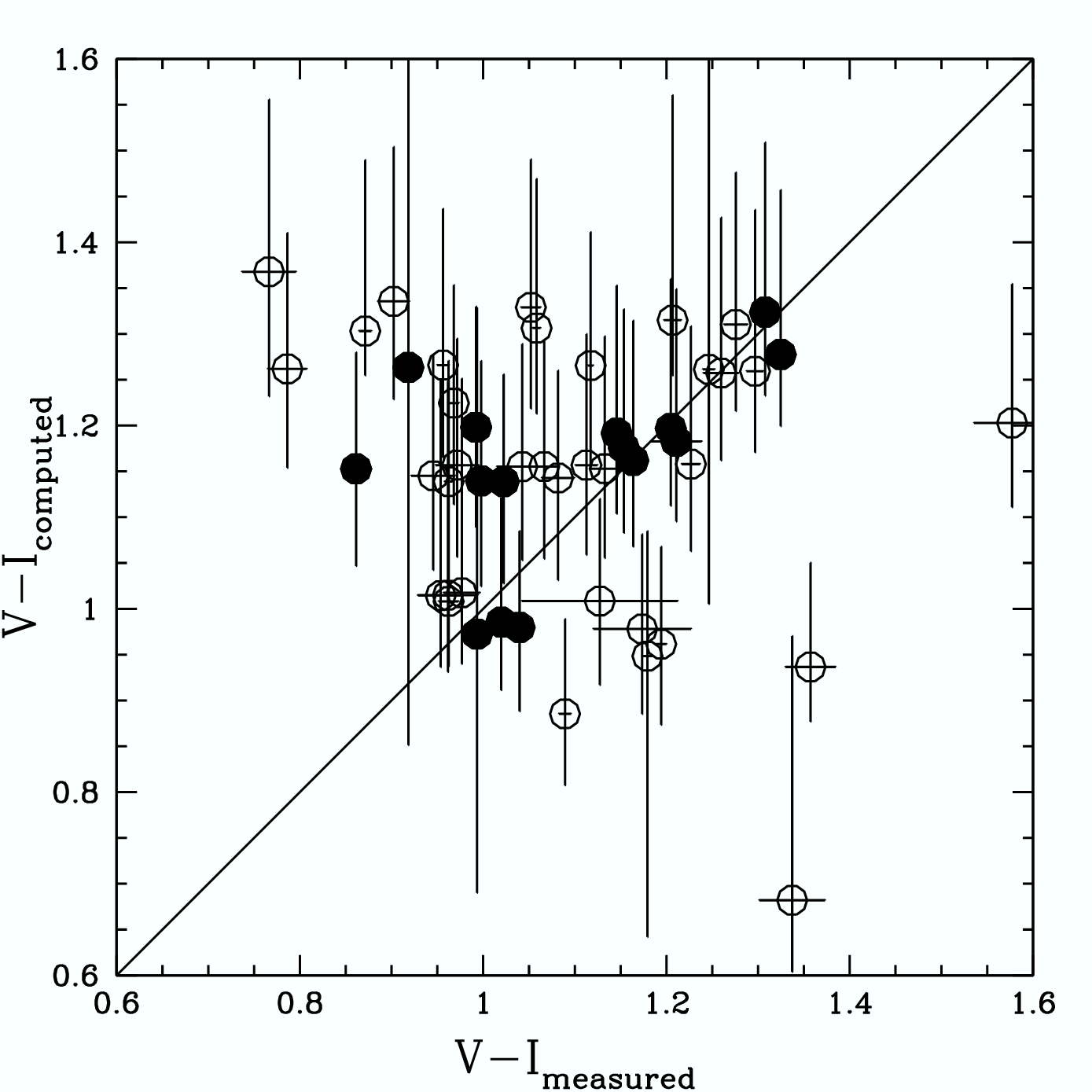}
\includegraphics[width=5.5cm,]{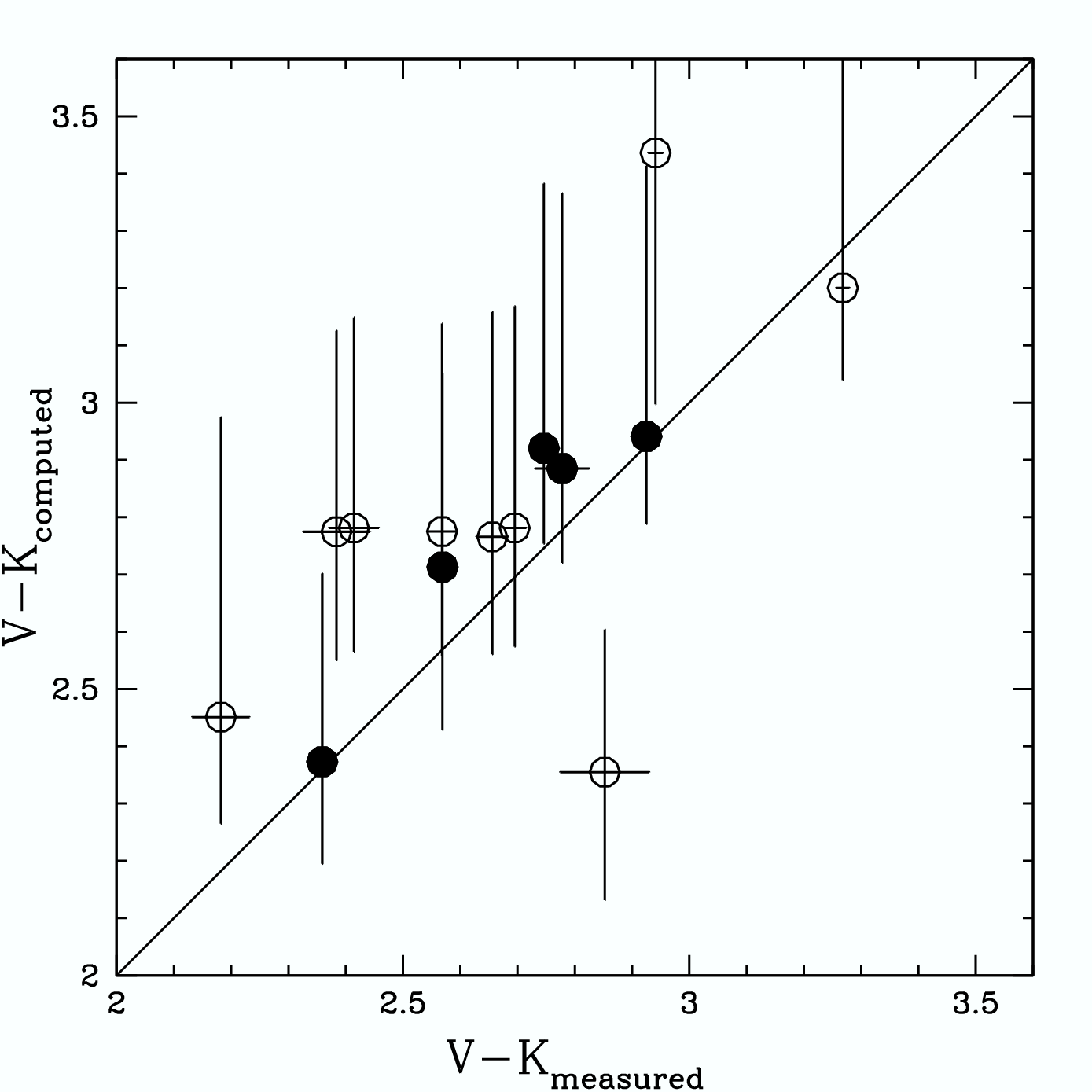}
 \caption{Computed vs. measured photometric colors for our sample
globular clusters. Computed colors were derived from spectroscopic ages
and metallicities using the SSP models of \cite{maraston03m}. The circles
show all selected globular clusters for which optical/near-infrared colors
were obtained (see Paper I for details). Solid circles show the
high-quality sample, which was selected by a more constrained error
cut (see text for details). Errorbars indicate 1-$\sigma$ uncertainties.}
 \label{ps:calccolors}
\end{figure*}

The comparison of derived formal ages and metallicities of individual
globular clusters shows that different diagnostic diagrams make {\it age}
predictions which are somewhat inconsistent with one another. The average
relative age uncertainty varies in the range $\Delta t/t\approx 0.2-0.5$
which we attribute to systematic errors in the age scale of model
predictions and/or systematics in the calibration of the data. By
combining age predictions from different diagnostic plots we smooth out
these systematic effects which influence age determinations derived from
single Balmer indices.

This age inconsistency is partly surprising as the Maraston et
al.~(2003) models are well calibrated for Milky Way globular clusters and
their ages derived from different Balmer indices are consistent with each
other. 

On the other hand, we find very good agreement of metallicity
predictions from all diagnostic plots within an average uncertainty of
$\sim0.2$ dex.

Note also, that a striking feature of all age/metallicity diagnostic plots
in Figure~\ref{ps:Balmer_Metal} is that a significant fraction of globular
clusters lie below the oldest iso-age track. This is observed in other
samples of {\it extragalactic} globular clusters in the literature as well
\citep[e.g.][]{kissler-patig98a, cohen98, puzia00, larsen03}. Systematics
errors in the data reduction have been deemed extremely unlikely given the
good Lick standard calibration (see Paper I) and are also unlikely given
the general problem in the literature. We could also exclude emission
filling as source for the Balmer index inconsistencies which is presented
in Section~\ref{ln:emission}.

To test whether the spectroscopic age and metallicity predictions for these
outliers are consistent with their observed photometric colors (see Paper
I for details), we use SSP models of \cite{maraston03m} to compute
photometric colors from the derived spectroscopic ages and metallicities.
Figure~\ref{ps:calccolors} shows the comparison of computed and observed
data. We find good agreement for the $B\!-\!V$,  $V\!-\!I$, and  $V\!-\!K$
colors. We find no systematics in color residuals as a function of Balmer
index with respect to the one-to-one relation. However, we find significantly smaller
residuals towards more metal-rich objects.

We also investigate if higher-quality data alleviates these
inconsistencies and select a sub-sample with statistical errors
$\Delta$H$\beta<0.11$ \AA\ and $\Delta$H$_{\gamma,\delta}<0.165$ \AA\ for
higher-order Balmer lines, which is shown as solid dots in
Figure~\ref{ps:calccolors}. For this sub-sample, we indeed find a better
agreement between computed and observed photometric colors. We take this
as an indication that the age and metallicity predictions for the
higher-quality dataset are more robust. Based on our photometric
cross-check, we only consider the ages and metallicities derived for our
high-quality sub-sample as trustworthy. However, to illustrate the
difference between the high-quality and the rest of the data we plot the
remaining dataset in the following diagnostic plots as open symbols in
Figure~\ref{ps:Balmer_Metal} and show their distributions as hatched
histograms in Figure~\ref{ps:amah}.

\subsection{Results}
\begin{figure*}[!ht]
\centering 
\includegraphics[width=6.2cm, bb=10 120 600 700]{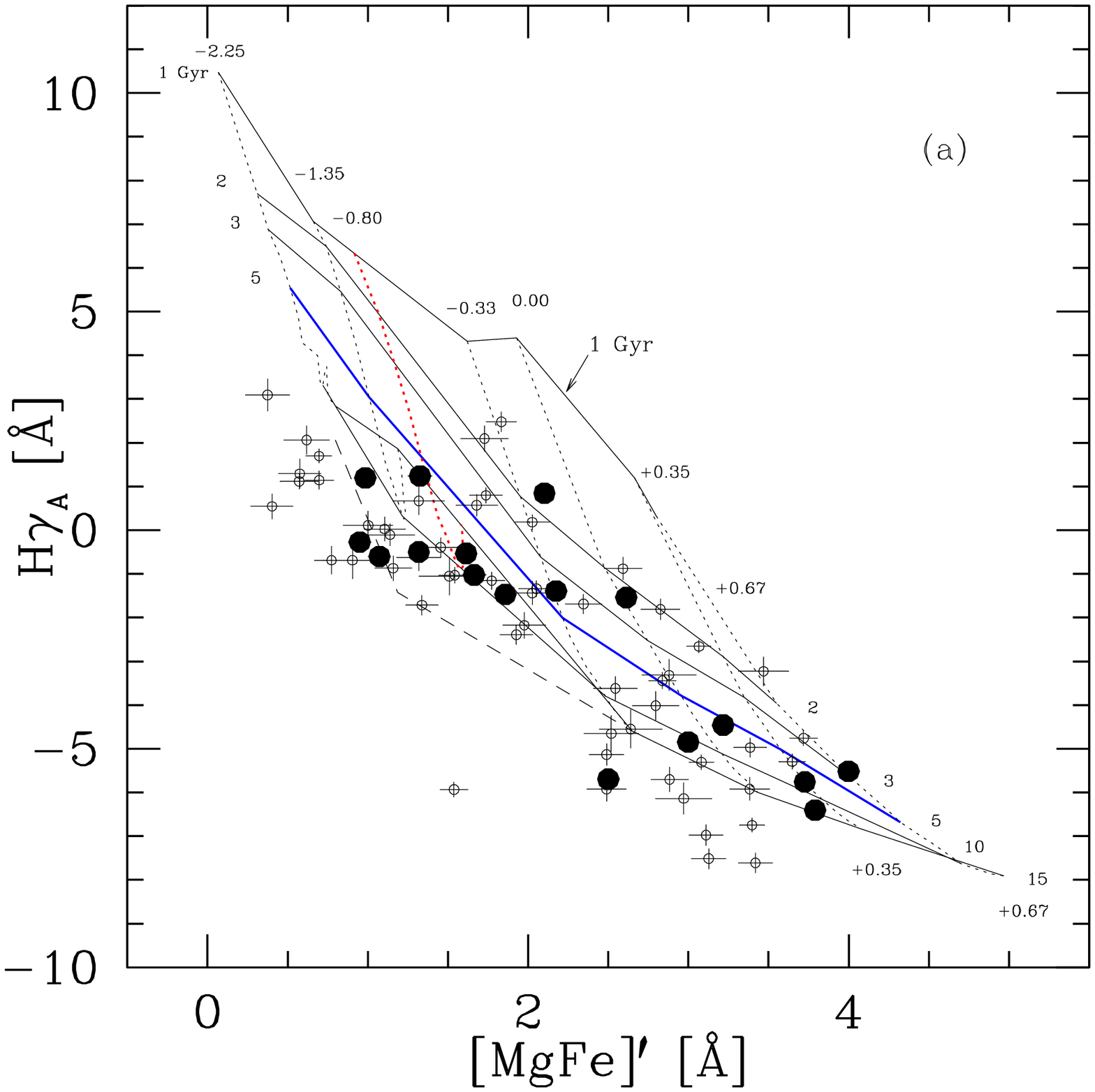}
\includegraphics[width=6.2cm, bb=10 120 600 700]{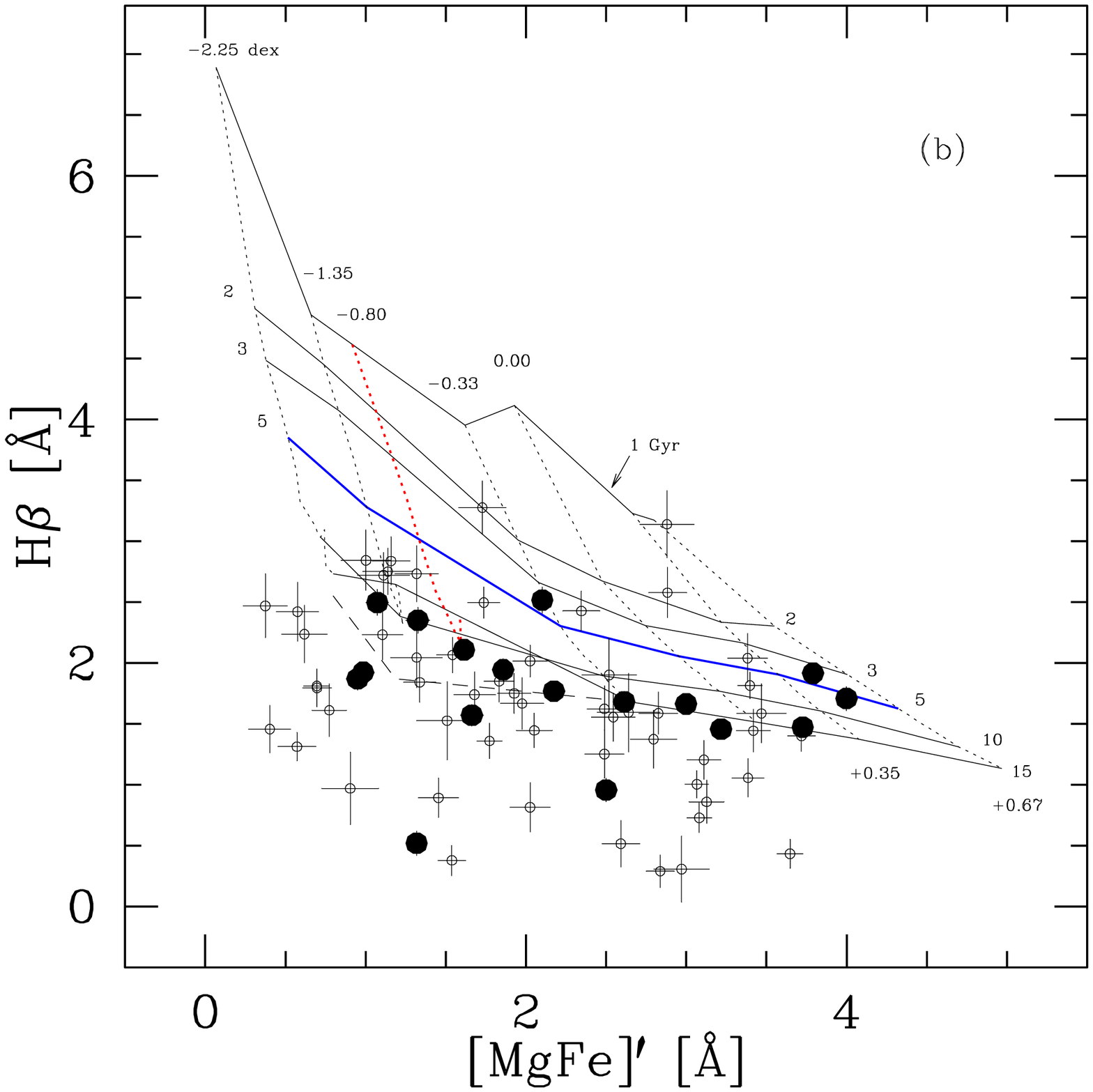}
\includegraphics[width=6.2cm, bb=10 120 600 700]{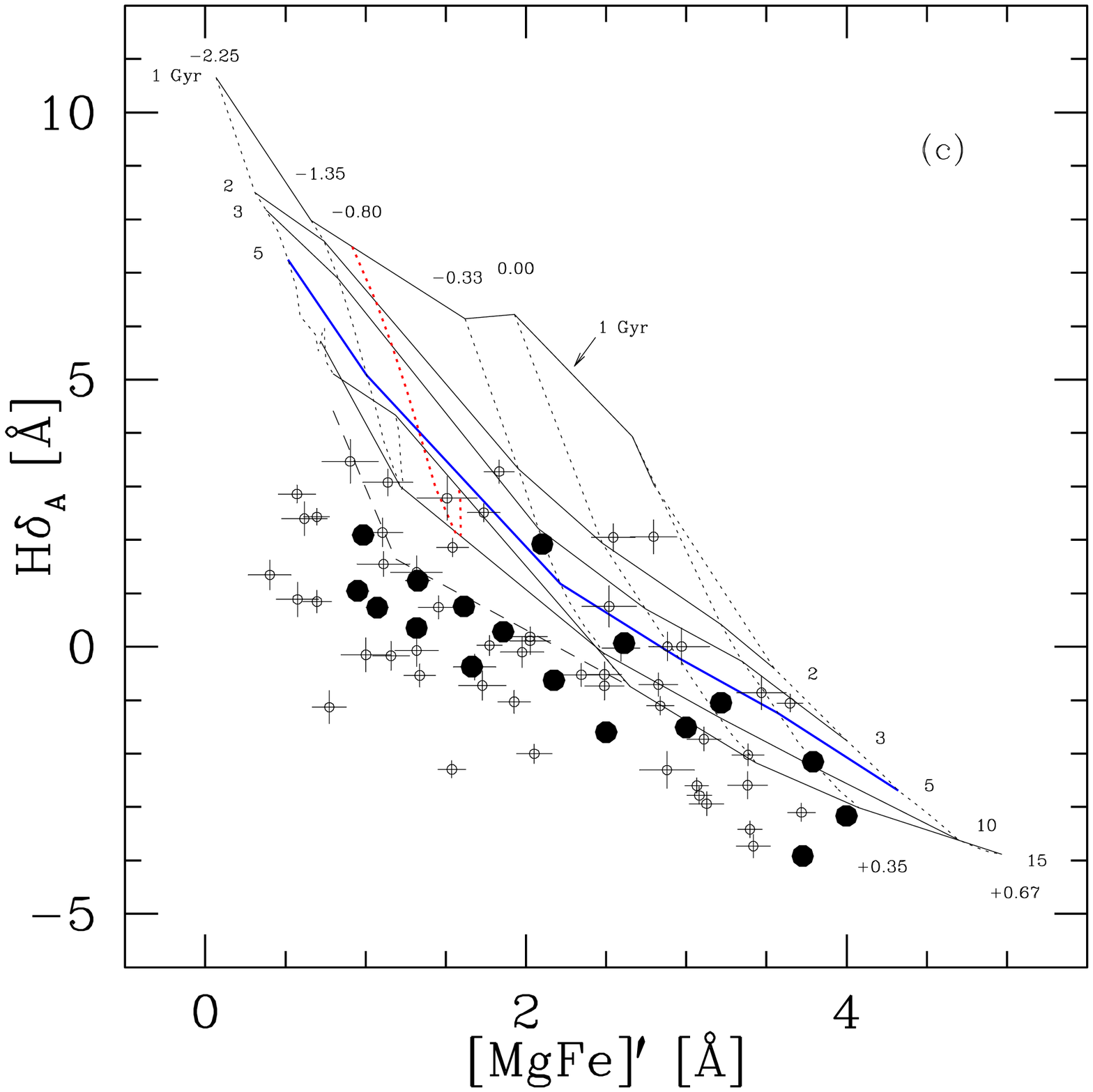}
\includegraphics[width=6.2cm, bb=10 120 600 700]{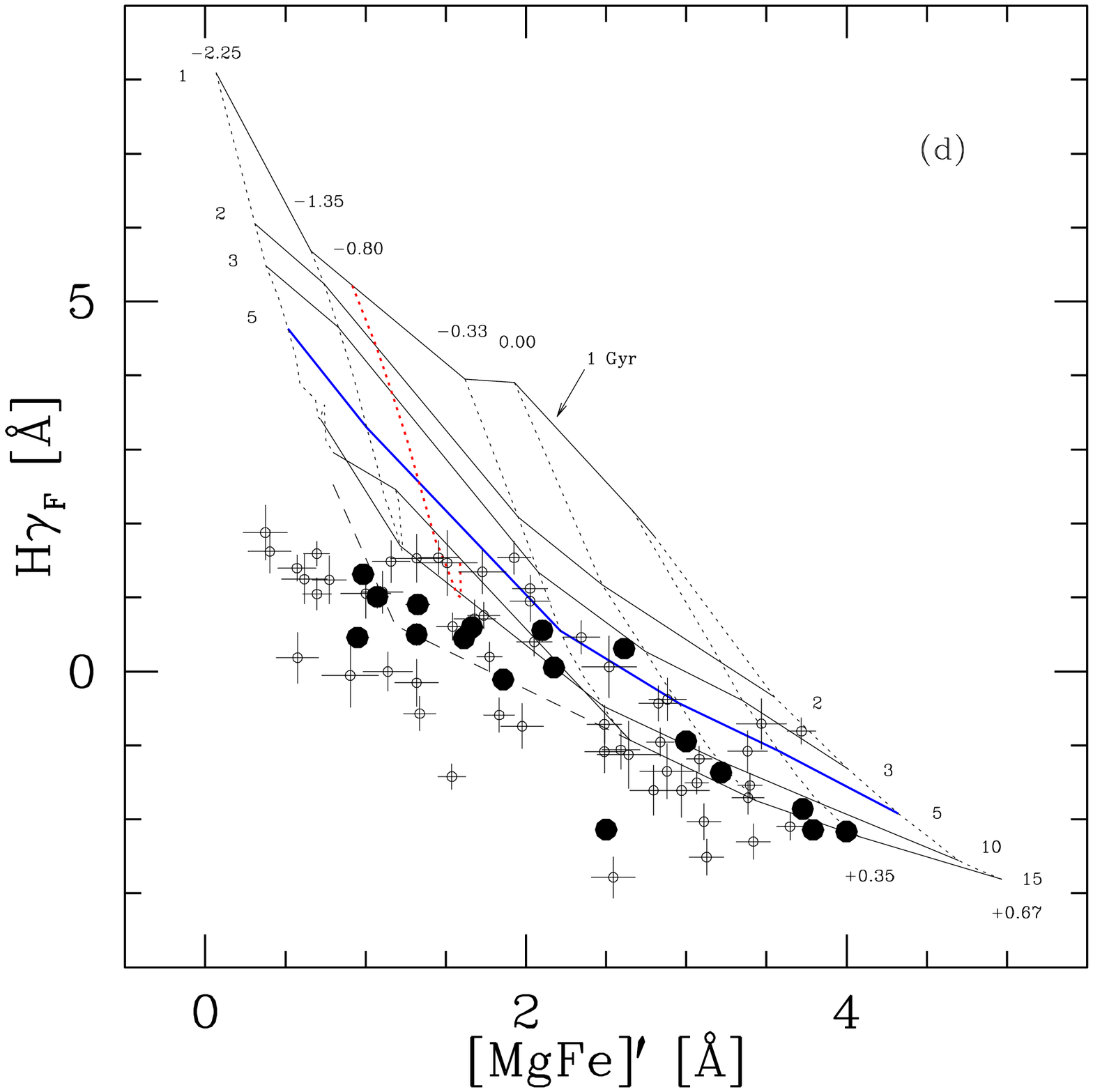}
\includegraphics[width=6.2cm, bb=10 120 600 700]{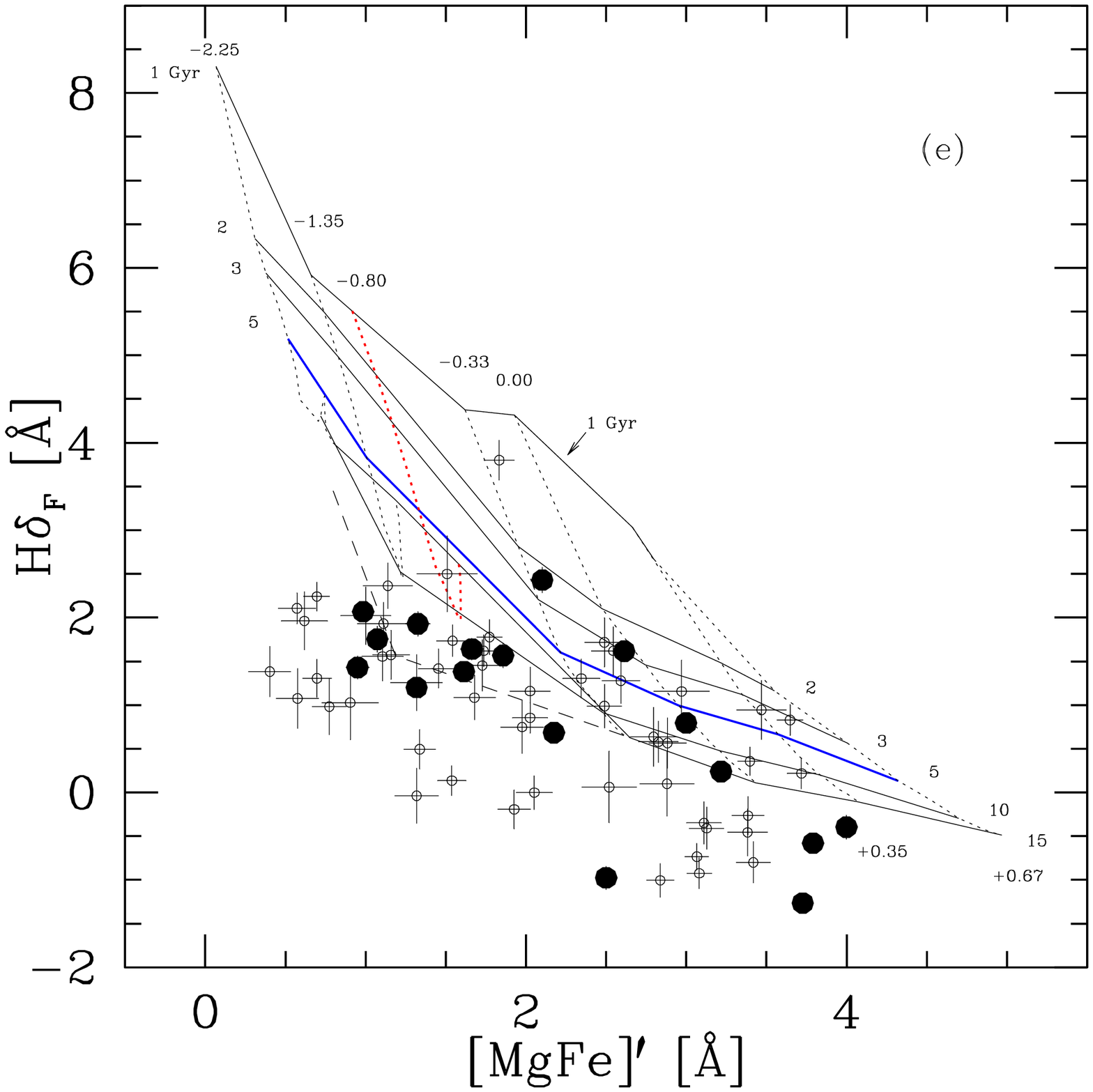}
\includegraphics[width=6.2cm, bb=10 120 600 700]{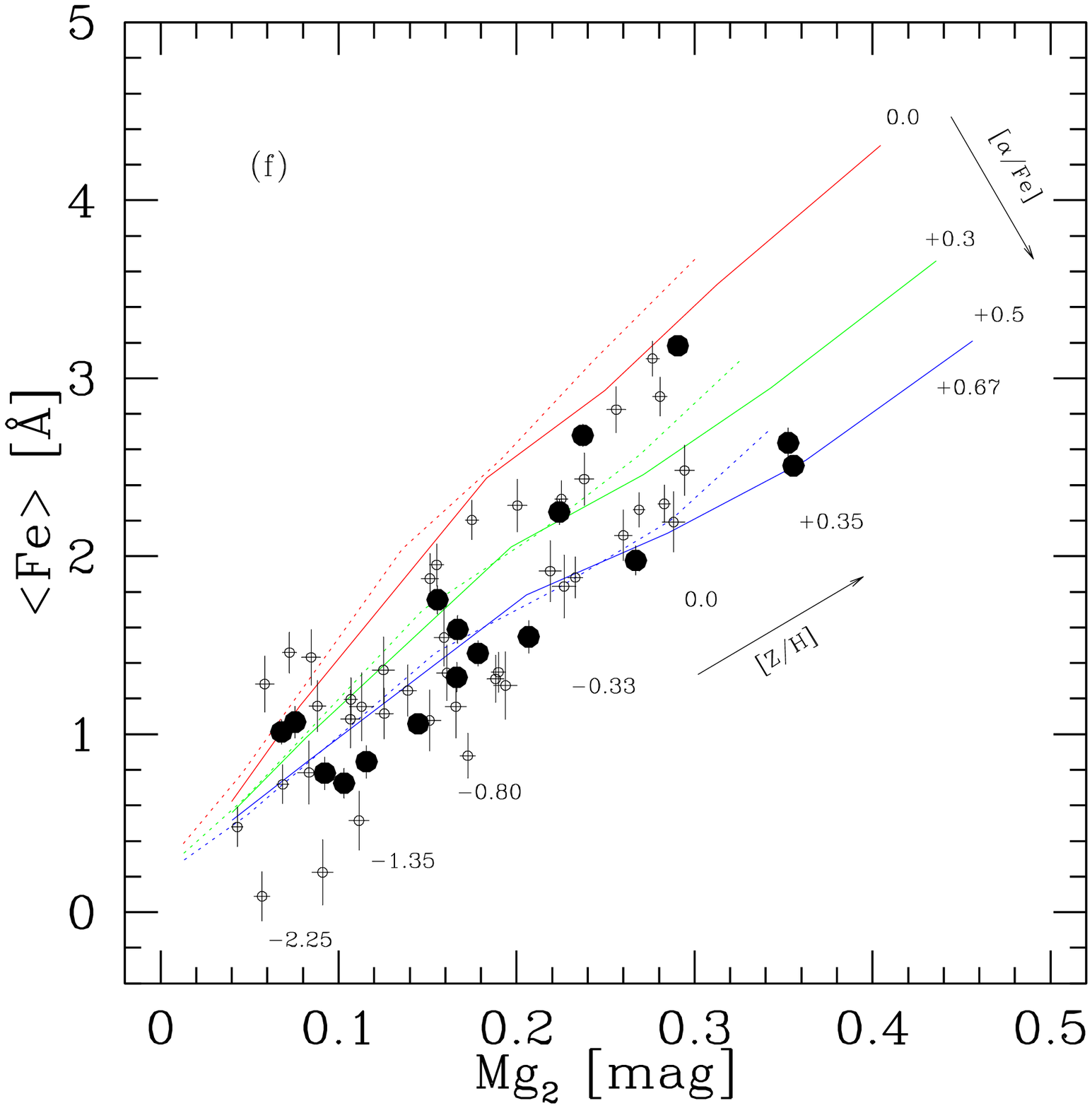}
\caption{Age-metallicity diagnostic plots ({\it panels a--e}) constructed from
Balmer indices H$\gamma_{\rm A}$, H$\beta$, H$\delta_{\rm A}$,
H$\gamma_{\rm F}$, and H$\delta_{\rm F}$ vs. [MgFe]\arcmin\ for
high-quality globular cluster spectra (solid dots). Open circles show our
remaining selected globular cluster data. SSP models of \cite{thomas04}
have been over-plotted for [$\alpha$/Fe]~$=0.3$, metallicities
[Z/H]=$-2.25, -1.35, -0.80, -0.33, 0.00$, 0.35, and 0.67 dex (dotted
lines), and ages 15, 10, 5, 3, 2, 1 (solid lines). The red dotted line
is an interpolated iso-metallicity track for [Z/H]~$=-0.8$ and is used to
split the sample between metal-poor and metal-rich globular clusters. The
blue solid line is the 5 Gyr isochrone, and is used to split between
old and young globular clusters. A {\it dashed isochrone} illustrates the Balmer
index strengths for a 15 Gyr old stellar population with an entirely red
horizontal-branch morphology. The [$\alpha$/Fe] diagnostic plot ({\it
panel f}) shows the mean iron index $\langle$Fe$\rangle$ as a function of
Mg$_2$. SSP models of \cite{thomas03, thomas04} with constant
[$\alpha$/Fe] ratios have been over-plotted for [Z/H] between $-2.25$ and
$+0.67$ dex and two ages 13 ({\it solid lines}) and 3 Gyr ({\it dotted
lines}) with various [$\alpha$/Fe] ratios 0.0, +0.3, and +0.5 dex.}
\label{ps:Balmer_Metal}
\end{figure*}

\begin{figure*}[!ht]
\centering
\includegraphics[width=14.0cm, bb=1 1 400 400]{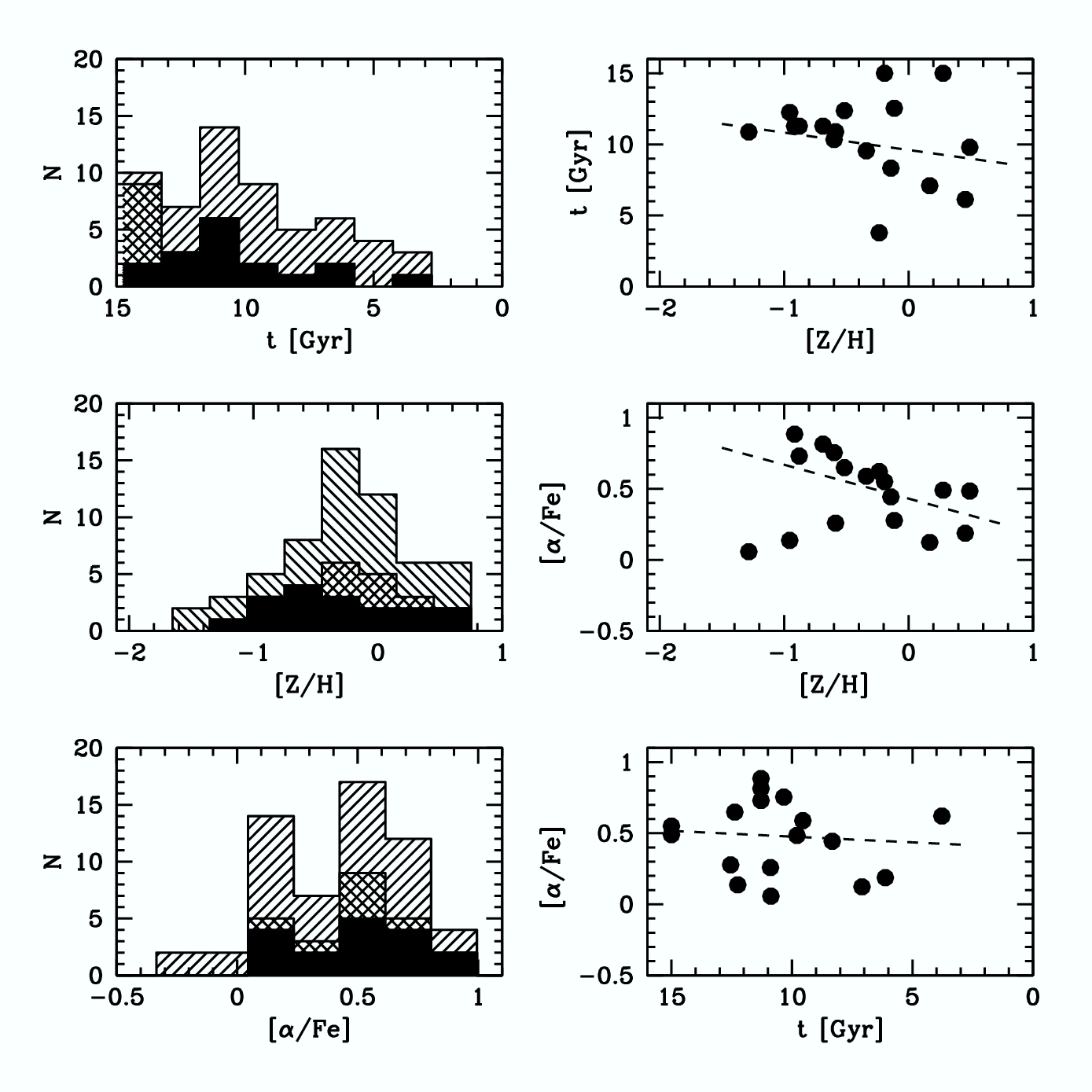}
   \caption{Histograms for ages, metallicities, and [$\alpha$/Fe] ratios 
for globular clusters in early-type galaxies. The panels in the left
column show distribution of each parameter for all selected (hatched
histograms), outliers in diagnostic plots (double-hatched histograms), and
high-quality data (solid histograms). Each panel in the right
column illustrates a correlation of two of these parameters for the
high-quality sub-sample. Dashed lines in the correlation plots show
weighted least-square fits.}
\label{ps:amah}
\end{figure*}

In the following, we discuss ages and metallicities for the high-quality
sub-sample derived from diagnostic plots using the Balmer-line indices
H$\beta$, H$\gamma_{\rm A}$, and H$\delta_{\rm A}$ (see
Sect~\ref{ln:conschecks}). The corresponding diagnostic diagrams are shown
in Figure~\ref{ps:Balmer_Metal} along with the other two diagrams
constructed from the Balmer indices H$\gamma_{\rm F}$, and H$\delta_{\rm
F}$. Since for seven globular clusters one or more of the three Balmer
indices H$\beta$, H$\gamma_{\rm A}$, and H$\delta_{\rm A}$ are not
available, we exclude these objects from the subsequent analysis to avoid
systematics, which leaves 17 globular clusters in our high-quality
sub-sample. Their age and metallicity distributions are shown in
Figure~\ref{ps:amah} as solid histograms.

Our method reveals that a significant fraction of clusters in the
high-quality sample (6/17 or 35\%, see solid histogram in
Fig.~\ref{ps:amah}) has formal ages between 5 and 10 Gyr. Only one object
has a formal age below 5 Gyr ($\sim3.8$ Gyr) and there are no globular
clusters with extremely young ages ($\la1$ Gyr). We split the sample at
[Z/H]~$=-0.6$, corresponding to the dip in the Milky Way globular cluster
metallicity distribution \citep{harris96}, into metal-poor and metal-rich
globular clusters. With this distinction we find an increase in age spread
from metal-poor to metal-rich clusters. We derive a dispersion of 0.5 Gyr
for metal-poor and a dispersion of 3.5 Gyr for metal-rich globular
clusters. A weighted linear least-square fit to our high-quality dataset
reveals a weak age-metallicity relation, in the sense that more metal-rich
globular clusters appear on average younger. The significance of this
relation needs to be tested with larger samples.

Our high-quality globular cluster sample covers metallicities in the range
$-1.3\la$~[Z/H]~$\la+0.5$, with a mean $-0.36\pm0.13$ and dispersion
$\sigma=0.52$ dex. Four out of 17 globular clusters have formally
super-solar metallicities. All clusters that are younger than 10 Gyr have
metallicities [Z/H]~$\ga-0.4$, which excludes their formation from
primordial gas clouds.

We note that no super metal-rich, inter\-mediate-age counterpart globular
cluster population is found in the Milky Way \citep[e.g.][]{harris01}.
However, globular clusters with intermediate ages are found in M31
\citep[e.g.][]{barmby00, burstein04, beasley04, puzia04}, whose
bulge-to-disk ratio is larger than that of the Milky Way. An interesting
idea is that this mode of globular cluster formation may not have
manifested itself in our Galaxy due to Milky Way's smaller bulge
\citep{goudfrooij03}.

In summary, the conclusions that can be drawn from our previous
analysis of globular clusters in early-type galaxies are the following:
\begin{itemize}
\item[$\bullet$] metal-poor globular clusters are on average older
  than metal-rich globular clusters
\item[$\bullet$] there is a tendency for metal-rich globular clusters 
  to have a younger mean age relative to their metal-poor counterparts
\item[$\bullet$] metal-rich globular clusters exhibit on average a
  larger age spread than metal-poor globular clusters
\end{itemize}

\section{[$\alpha$/Fe] Ratios}
\label{ln:alpha}
In this Section we derive [$\alpha$/Fe] ratios for our high-quality sample globular
clusters using a diagnostic diagram which is least sensitive to
age/metallicity variations. Such a diagram can be constructed from the
indices $\langle$Fe$\rangle$ and Mg$_2$, which primarily trace the
abundances of iron and the $\alpha$-element magnesium \citep{tripicco95}.
We note that among the three Mg-sensitive indices, Mg$_1$, Mg$_2$, and
Mg$b$, theoretical index predictions show a
relatively large spread in $\langle$Fe$\rangle$ and Mg$_2$ for
[$\alpha$/Fe] ratios between solar and $\sim0.5$ dex, at high mean
metallicities. Given the quality of our data we can expect a good
discrimination between enhanced and solar-type [$\alpha$/Fe] ratio for
individual globular clusters at metallicities [Z/H]~$\ga-0.8$ and a good
estimate of the mean [$\alpha$/Fe] ratio at lower metallicities.

Figure~\ref{ps:Balmer_Metal} (panel $f$) shows that the Mg$_2$
vs. $\langle$Fe$\rangle$ diagnostic diagram is not entirely free from the
age/metallicity degeneracy. Iso-[$\alpha$/Fe] tracks for three different
ratios (0.0, 0.3, and 0.5 dex) are plotted for two ages (3 and 13 Gyr,
indicated by dotted and solid lines, respectively). It is obvious that age
information is needed to choose the correct set of tracks for a reliable
[$\alpha$/Fe] determination. The diagnostic grid is interpolated in our
iterative fitting routine using information derived from age/metallicity
diagnostic grids (see Sect.~\ref{ln:method}). In other words,
[$\alpha$/Fe] ratios are determined simultaneously with ages and
metallicities. The averaging is identical to the one for ages and
metallicities in Section~\ref{ln:method}.

A histogram of [$\alpha$/Fe] ratios for our high-quality sample is shown
in Figure~\ref{ps:amah}. {\it All} globular clusters are consistent with
super-solar [$\alpha$/Fe] ratios. The mean [$\alpha$/Fe] of the sample is
$0.47\pm0.06$ dex, with a dispersion of $0.26$ dex.

Using a weighted linear least-square fit, we find evidence for a
[$\alpha$/Fe]--metallicity relation in the sense that more metal-rich
globular clusters have lower [$\alpha$/Fe] ratios. However, due to the
reduced [$\alpha$/Fe] resolution of Lick indices at low metallicities and
our modest sample size (especially at the metal-poor end), this trend
needs to be confirmed with more data. Consistent with the previously found
age--metallicity and [$\alpha$/Fe]--metallicity relation, we find no
evidence for a [$\alpha$/Fe]--age correlation. Globular clusters at all
ages appear to have on average super-solar [$\alpha$/Fe] ratios.

Recent measurements of [$\alpha$/Fe] ratios in globular cluster systems in
other early-type galaxies reveal very similar results. For instance, the
data of \cite{kuntschner02b} for globular clusters in NGC~3115 show that
most clusters are consistent with [$\alpha$/Fe]~$\approx0.3$ over the
entire range of sampled metallicities. \cite{larsen02b} find a super-solar
mean [$\alpha$/Fe] ratio of $+0.4$ dex for globular clusters of all
metallicities in NGC~4594 (Sombrero). Using SSP models with non-constant
[$\alpha$/Fe] ratios, \cite{forbes01} argue that at least some globular
clusters in NGC~1399 exhibit super-solar [$\alpha$/Fe] ratios. However,
the found [$\alpha$/Fe]--metallicity correlation is seen in this sample for
the first time. This is mainly due to the higher quality of our data and a
more reliable age/metallicity determination compared to previous studies.

Note that super-solar [$\alpha$/Fe] ratios have also been reported for the
{\it diffuse light} of early-type galaxies \citep[e.g.][]{davies93,
trager00b, davies01, thomas02, kuntschner02a}. In subsequent papers of
this series we will compare the index measurements of globular clusters
with indices derived from our long-slit spectra of the hosts' diffuse
stellar light.

In the following we summarize the major findings of this section:
\begin{itemize}
\item[$\bullet$] globular clusters in early-type galaxies have on
  average super-solar [$\alpha$/Fe] ratios with a mean $\sim0.45$ dex
\item[$\bullet$] metal-rich globular clusters have on average lower
  [$\alpha$/Fe] ratios than their metal-poor counterparts
\item[$\bullet$] young and old globular clusters have on average
  similar [$\alpha$/Fe] ratios
\end{itemize}

\section{Discussion}
\label{ln:discussion}
\subsection{Assembly History of Early-Type Galaxies}

\begin{figure*}[!th]
\centering
\includegraphics[width=14cm, bb=1 1 400 400]{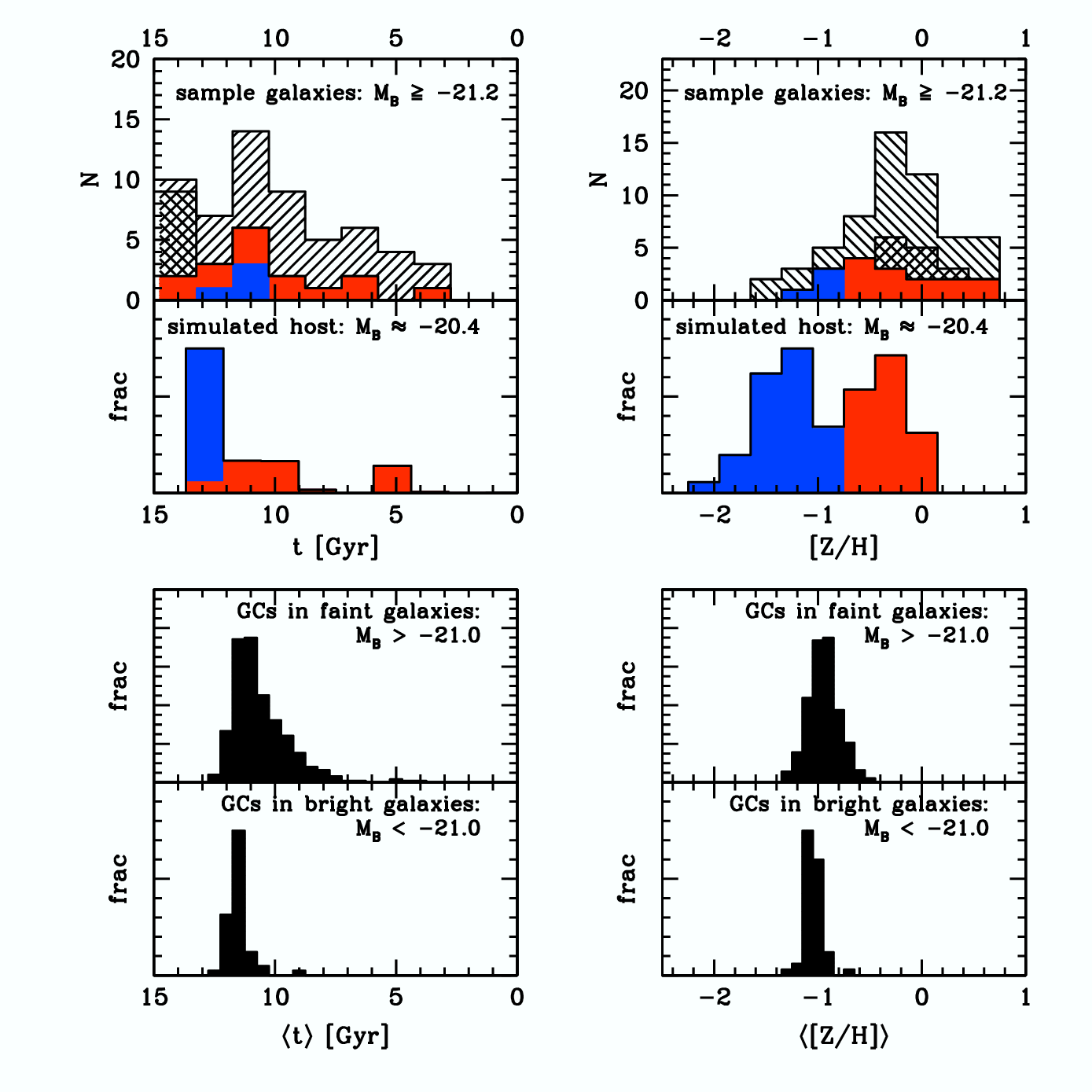}
   \caption{{\it Top row}: Comparison of globular cluster ages and
metallicities in our sample galaxies with those in a simulated host galaxy
with a similar luminosity (mass). All globular cluster in our sample are
shown as hatched histograms, outliers in diagnostic plots are indicated by
double-hatched histograms (see Sec.~\ref{ln:conschecks}), and high-quality
data is shown as solid histograms. Observed and simulated globular cluster
samples were split into metal-poor and metal-rich globular clusters at
[Z/H]~$=-0.9$ and the shading is accordingly transfered to the age
distribution plot. {\it Bottom row}: Average ages and metallicities of
simulated globular cluster systems in faint and bright host galaxies. All
predictions were taken from \cite{beasley02a}.}
\label{ps:cdmcomp}
\end{figure*}

If the dispersion in Balmer-[MgFe]\arcmin\ diagnostic plots is entirely
driven by age\footnote{A population of warm horizontal branch stars
remains a potential cause for increased Balmer line strengths.}, we find
indications for a significant fraction of relatively young globular
clusters in early-type galaxies. About $\sim1/3$ of our sample globular
clusters have formal ages younger than $10$ Gyr, implying formation
redshifts $z_f\la 1.7$ (in a $\Lambda$CDM {\it concordance} universe).

In Figure~\ref{ps:cdmcomp} we compare the derived ages and metallicities
of our high-quality sample with the corresponding distributions of
globular cluster systems simulated according to a hierarchical merging
scenario \citep{beasley02a}. As the average luminosity of our sample host
galaxies is $\langle M_B\rangle_{\rm sample}=-20.4\pm0.6$ (see
Tab.~\ref{tab:galdat}), we use the globular cluster system of a typical
simulated galaxy with a similar luminosity $M_{B}=-20.42$ \citep[see also
Fig.~6 in][]{beasley02a} for comparison. We split both cluster samples
(simulated and observed) in metal-poor and metal-rich clusters at
[Z/H]~$=-0.9$. The top right panel of Figure~\ref{ps:cdmcomp} shows that
our sample is biased towards metal-rich globular clusters, and the
comparison of metal-poor globular clusters is limited to the statement
that both observed and simulated clusters are all consistently old. For
metal-rich globular clusters there is good agreement in the age range
covered by data and simulations.

The somewhat naive prediction of more extended formation timescales
of more massive structures in the hierarchical picture holds only if the
fraction of gas-poor to gas-rich mergers, the so-called dry to mixed
merger ratio, is constant with redshift. \cite{khochfar03} predict that
this ratio depends on galaxy mass in the sense that most massive
ellipticals formed early in rather dissipationless (dry) mergers of
bulge-dominated precursors \citep[see also][]{kauffmann00}. This would
imply that on average low-luminosity ellipticals would harbor a higher
fraction of young globular clusters, while most massive galaxies would
preferentially host old globular cluster systems. This is reflected in the
simulations of \cite{beasley02a}. The lower panels of
Figure~\ref{ps:cdmcomp} illustrate the mean ages and metallicities of
globular cluster systems in faint and bright galaxies with a cut at
$M_{B}=-21.0$. The plots show that while host galaxy luminosity affects
the mean metallicity of globular cluster systems only marginally, a
significant fraction ($\sim25$\%) of low-luminosity galaxies is predicted
to host globular cluster systems with an average age younger than $\sim10$
Gyr, compared to a tiny fraction ($\sim2$\%) in bright galaxies.
Unfortunately, the transition between dry and mixed merger-dominated
evolution is predicted to occur for early-type galaxies in the luminosity
range between $M_B\approx-20$ and $-21$ mag, where all of our sample
galaxies reside.

To first order, previous spectroscopic studies find evidence for mostly
old globular cluster systems in the massive Fornax galaxy NGC~1399
\citep{kissler-patig98a}, and the two Virgo galaxies M87 \citep{cohen98}
and M49 \citep{cohen03}, which appears to fit into the framework of {\it
dissipationless} merging with an early assembly. We will get back to this point
in a subsequent paper in which globular cluster systems will be
investigated as a function of the host galaxy property.

\subsection{Formation Timescales}
Stellar populations with super-solar [$\alpha$/Fe] ratios, as they are
observed in massive elliptical galaxies, are generally interpreted as the
result of very short formation time scales. However, in the case of
dissipative merging, hierarchical merging models predict frequent merging
events, which is expected to result in relatively low [$\alpha$/Fe] ratios
\citep[e.g.][]{thomas99}. Although somewhat on the high side, our formal
mean [$\alpha$/Fe] ratio of $\sim0.47$ dex for globular clusters in
early-type galaxies is in line with values measured for the diffuse light.
Hence, the formation time scales of field stars and globular clusters
appear to be similar in elliptical galaxies and not immediately compatible
with hierarchical merging models.

The found evidence for a [$\alpha$/Fe]--metallicity relation in early-type
globular cluster systems is consistent with a generic chemical enrichment
scenario in which more metal-rich stellar populations have lower
$\alpha$-element enhancements. Two types of supernovae contribute to the
build-up of $\alpha$-peak (type II) and iron-peak elements (type Ia) in
the interstellar medium on different time scales
\citep[e.g.][]{matteucci94, thomas99}, because of the evolutionary delay
of $\sim1$ Gyr of type-Ia supernova progenitors \citep{greggio97}. Since
globular clusters are not massive enough to support significant
self-enrichment, their [$\alpha$/Fe] ratios reflect the {\it large-scale}
chemical conditions during their formation. However, to allow a detectable
increase in [$\alpha$/Fe], for high-metallicity stellar populations it
requires a larger number of type II supernovae to outweigh any previous
metal-enrichment by type Ia supernovae compared to lower metallicities.
This implies that high [$\alpha$/Fe] ratios at high metallicities are
likely the result of {\it shorter} formation timescales and/or higher
star-formation rates than similar [$\alpha$/Fe] ratios at lower
metallicities. Careful modeling of the chemical evolution of a globular
cluster system is needed to quantify these formation timescales. The
relation between [$\alpha$/Fe] and metallicity found here provides the
first important constraint for this exercise.

It is also important to test whether a steeper [$\alpha$/Fe]--age relation
can be found for globular cluster systems in more massive galaxies, which
would be expected if these systems formed on more extended timescales, as
predicted by hierarchical merging models. 

\section{Summary and Conclusions}
We have conducted a study of ages, metallicities, and [$\alpha$/Fe] ratios
of extragalactic globular clusters in early-type galaxies, based on the
Lick index system. We find that up to $\sim\!1/3$ of our sample of globular
clusters have ages formally younger than 10 Gyr. This result is not
biased by one single globular cluster system in our galaxy sample and
appears representative for early-type galaxies in general. We cannot state
with confidence whether the younger ages (i.e. $t\!<\!10$ Gyr) are real or
due to an unexpected blue horizontal branch morphology at high
metallicities. If the younger ages are real, the found fraction should be
taken as an upper limit, since our data only probe the bright end of the
globular cluster luminosity function where relatively young globular
clusters, if present, are expected to reside. In general, less than
$\sim\!1/3$ of the brightest $\sim\!10$\% of globular cluster systems in
early-type galaxies which we sample with our study could have formed at
redshifts $z_f\la1.7$.

We find that the formal age scatter increases and the mean age decreases
from metal-poor to metal-rich globular clusters, resulting in an
age--metallicity relation. For our high-quality sample globular clusters
with metallicities [Z/H]~$<-0.6$ we find an age dispersion of 0.5 Gyr. The
metal-rich sub-sample has a dispersion of 3.5 Gyr.

Our high-quality sample spans a wide range in metallicity between
[Z/H]~$\approx-1.3$ and $\sim+0.5$ dex. We find evidence for an
[$\alpha$/Fe]--metallicity relation in the sense that more metal-rich
globular clusters have lower $\alpha$-element enhancements, which needs to
be confirmed with more data that better sample the metal-poor regime.
However, there is no indication for an [$\alpha$/Fe]--age correlation.

[$\alpha$/Fe] ratios are found to be on average super-solar with a mean
$0.47\pm0.06$ dex and a dispersion of $\sim\!0.3$ dex, which indicates
formation timescales shorter than $\sim\!1$ Gyr. In other words, the
progenitor clouds of globular clusters were predominantly enriched by
type-II SNe.

\begin{acknowledgements}
We are grateful to Michael Rich for providing colour magnitude diagrams of
M31 globular clusters prior to publication. Many thanks go to Michael
Beasley for sending electronic tables of his hierarchical-clustering
simulations. We also thank Scott Trager for a very constructive referee
report. THP gratefully acknowledges the support by the German
\emph{Deut\-sche For\-schungs\-ge\-mein\-schaft, DFG\/} project number
Be~1091/10--2, and the support in form of an ESA Research Fellowship.
\end{acknowledgements}

\bibliographystyle{apj}

\end{document}